\documentclass[prb, aps, bibliography, twocolumn, reprint, showpacs, footnoteinbib]{revtex4-1}
\usepackage{graphicx}
\usepackage{amssymb}   
\usepackage{amsfonts}
\usepackage{amsmath}
\usepackage{dsfont}
\usepackage{dcolumn}   
\usepackage{bm}        
\usepackage[mathscr]{eucal}
\usepackage[dvipsnames]{xcolor}
\usepackage[colorlinks, linkcolor=OrangeRed,citecolor=RoyalBlue,urlcolor=NavyBlue]{hyperref}
\usepackage[all]{hypcap} 
\usepackage{mathtools}
\usepackage{wasysym}

\newcommand{\B} {\bf}

\begin{document}

\title{Constructing effective free energies for dynamical quantum phase transitions in the transverse-field Ising chain}
\author{Daniele Trapin}
\author{Markus Heyl}

\affiliation{Max-Planck-Institut f\"ur Physik komplexer Systeme, N\"othnitzer Stra{\ss}e 38,  01187-Dresden, Germany}

\begin{abstract}
The theory of dynamical quantum phase transitions represents an attempt to extend the concept of phase transitions to the far from equilibrium regime.
While there are many formal analogies to conventional transitions, it is a
major question to which extent it is possible to formulate a nonequilibrium
counterpart to a Landau-Ginzburg theory.
In this work we take a first step in this direction by constructing an effective free energy for continuous dynamical quantum phase transitions appearing after quantum quenches in the transverse-field Ising chain. 
Due to unitarity of quantum time evolution this effective free energy becomes a complex quantity transforming the conventional minimization principle of the free energy into a saddle-point equation in the complex plane of the order parameter, which as in equilibrium is the magnetization.
We study this effective free energy in the vicinity of the dynamical quantum phase transition by performing an expansion in terms of the complex magnetization and discuss the connections to the equilibrium case.
Furthermore, we study the influence of perturbations and signatures of these dynamical quantum phase transitions in spin correlation functions.
\end{abstract}

\maketitle
\section{Introduction} \label{Introduction}
The theory of phase transitions plays a key role for the understanding of equilibrium matter. Of particular importance are continuous phase transitions, which exhibit scaling and universality implying that macroscopic properties become independent of
microscopic details. A unifying description for these transitions
is provided by Landau-Ginzburg theory~\cite{sachdev2007,huang2009}.
In the out-of-equilibrium regime, a temporal analogue of phase transitions has been introduced, termed dynamical quantum phase transitions 
(DQPTs)~\cite{heyl2013,Zvyagin2017Review,Heyl2017Review}. These DQPTs are driven not by a control parameter but rather occur 
as a function of time. This leads to a nonanalytic temporal behavior in physical quantities, which has recently been observed experimentally~\cite{Flaeschner2017, jurcevic2016}. While it has been shown that many properties of equilibrium transitions such 
as robustness~\cite{ScalingHeyl,sharma2015quenches,kriel2014dynamical,karrasch2013dynamical}, dynamical order parameters~\cite{budich2016dynamical,sharma2015quenches,Flaeschner2017,bhattacharya2017mixed,HeylBudich2017} or scaling and universality~\cite{ScalingHeyl} also apply to DQPTs, it remains a challenging problem to understand to which extend other central concepts of conventional phase transitions such as Landau-Ginzburg theories can also be formulated for DQPTs.

In this work, we construct a dynamical analogue to free energies for continuous DQPTs occuring in the nonequilibrium dynamics of the transverse-field Ising chain.
The key element of our construction is to map the central object of DQPTs, the Loschmidt amplitude,
onto a conventional classical Ising partition function with complex couplings for a certain quantum quench~\cite{ScalingHeyl,karrasch2017}.
In this way, we can utilize methods and concepts for the equilibrium Ising model upon generalizing to complex parameters, which straightforwardly allows us to formulate the anticipated effective free energy.
The unitarity of quantum real-time evolution makes the effective free energy a complex quantity.
As a consequence, the conventional minimization principle turns into a saddle-point condition in the plane of the complex order parameter.
We find that in the vicinity of the saddle point the effective free energy admits an expansion in powers of the complex magnetization, which plays the role of the order parameter, analogous to what one would obtain in the spirit of a Landau theory.
Since our model is exactly solvable, we can determine the expansion parameters from first principles.
We compare our findings for the effective free energy to the equilibrium free energy of the classical Ising chain.

The mapping of Loschmidt amplitudes to classical partition function makes it also possible to formulate exact renormalization group transformations which allows to identify the exact fixed points and therefore the nature of the DQPTs, which are known to be of continuous type for the considered nonequilibrium setup~\cite{ScalingHeyl}.
In order to further investigate similarities of DQPTs to conventional phase transitions, we study the influence of various perturbations to our model via their behavior under RG transformations. 
We find that those weak perturbations which preserve the $\mathbb{Z}_2$ symmetry of the transverse-field Ising chain, turn out to be irrelevant in the RG sense, which for some cases was already known before~\cite{heyl2013}.
Upon adding a longitudinal field in ordering direction, which breaks the Ising $\mathbb{Z}_2$ symmetry, we find that the nature of the DQPT changes from  continuous to first order.

Finally we connect the appearance of DQPTs in the Loschmidt amplitude to the dynamics of a local observable.
Since the DQPT is associated with an unstable fixed point and thus with a divergent correlation length, it is natural to expect that the dynamics of spin-spin correlations is strongly influenced by the underlying DQPT.
We find that spin-spin correlations show a marked signature of the DQPTs in that they become maximal whenever a DQPT occurs, as observed also in other contexts~\cite{ScalingHeyl,schmitt2017quantum}.

This paper is organized as follows.
We start by giving an introduction to the theory of DQPTs in Sec.~\ref{RG DQPTs} and an introduction to our studied model system, the transverse-field 
Ising chain, in Sec.~\ref{model}. Afterwards we briefly summarize our main results in Sec.~\ref{sec:mainresults} and then in Sec.~\ref{DQPTs} we explain how to map
the Loschmidt amplitude onto a classical partition function.
After introducing the effective free energy in Sec.~\ref{Dynamical Free Energy} , we use the complex magnetization to perform an expansion of the effective free energy in the vicinity of the DQPT in the spirit of Landau theory in Sec.~\ref{Landau}.
Finally, we move away from the infinite quench taking into account perturbations both in the initial and in the final Hamiltonians.
In Sec. \ref{SBP} we consider symmetry-breaking perturbations and their influence on DQPTs, while in Sec. \ref{SPP} we introduce a symmetry-preserving perturbation.

\section{Dynamical Quantum Phase Transitions} \label{RG DQPTs}
In the following we are interested in a genuine nonequilibrium quantum regime where the state of the system cannot be captured within equilibrium statistical physics in the sense that it cannot be described through a partition function or conventional free energy.
Still systems can undergo a dynamical quantum phase transition (DQPT) with physical quantities showing nonanalytic properties as a function of time~\cite{heyl2013,Heyl2017Review,Zvyagin2017Review}.
In this work, we study such nonanalytic temporal structures in the nonequilibrium dynamics generated by so-called quantum quenches~\cite{RevModPhys.83}.
In that context, the system is initially prepared in the ground state $|\psi\rangle$
of a Hamiltonian $H_0$. At time $t=0$ the system suddenly undergoes
a change of its parameters such that the dynamics is afterwards driven by a final Hamiltonian $H$. Solving formally the Schr\"odinger equation gives for the state $|\psi(t)\rangle$ at a time $t$:
\begin{equation}
	|\psi(t) \rangle = e^{-iHt} |\psi\rangle \, .
\end{equation}
For the theory of DQPTs after such quantum quenches, a central role is played by the Loschmidt amplitude:
\begin{equation}
 \mathcal{G}(t) = \langle \psi | \psi(t)\rangle =  \langle \psi | e^{-itH}|\psi\rangle,
 \label{G}
\end{equation}
which is the overlap between the initial $(|\psi\rangle)$ and the time-evolved state $(|\psi(t)\rangle)$, respectively. Consequently,
$\mathcal{G}(t)$ quantifies the deviation from the initial condition. Due to the formal similarities of $\mathcal{G}(t)$ to equilibrium
partition functions $Z(\beta)$, but even more to so-called boundary partition functions~\cite{leclair1995boundary}, it is suitable to define a dynamical analog of a free energy density (up to normalization) as:
\begin{equation}
 f(t) = -\lim_{N \rightarrow \infty} \frac{1}{N} \text{log}(\mathcal{G}(t)),
 \label{f}
\end{equation}
with $N$ denoting the number of degrees of freedom.
As conventional free energies can exhibit nonanalytic behavior at phase transitions, so can $f(t)$ in the thermodynamic limit.
In analogy to equilibrium, we define a DQPT as a point in time where the effective free energy $f(t)$ shows a nonanalytic structure.
While equilibrium phase transitions occur by varying an external parameter such as temperature controllable from the exterior, at a DQPT the system experiences nonanalytic behavior triggered solely by its intrinsic dynamics.

The analogies between DQPTs and conventional phase transitions are not limited to this formal level, but rather extend further.
This includes scaling and universality at continuous phase transitions where macroscopic properties become independent of microscopic details.
It has been shown that these concepts can also be applied to  DQPTs in the 1D Ising chain~\cite{ScalingHeyl}.
Moreover, DQPTs appear to be robust against small symmetry-preserving perturbations, meaning
that their presence cannot make  DQPTs disappear but they only contribute with some minor effects such as a shift in the critical time~\cite{karrasch2013dynamical,kriel2014dynamical,sharma2015quenches}.
Also dynamical order parameters have been identified~\cite{budich2016dynamical,sharma2016slow,bhattacharua2017a,Flaeschner2017,bhattacharya2017mixed,HeylBudich2017} which have been successful in characterizing DQPTs occurring in topological systems~\cite{vajna2015topological,schmitt2015dynamical,huang2016,Bhattacharya2017i,mera2017dynamical}.

\section{ Model and nonequilibrium protocol} \label{model}
In this work we study the one-dimensional (1D) Ising model in a transverse field:
\begin{equation}
	 H = -J \sum_n \sigma_n^z \sigma_{n+1}^z - h \sum_n \sigma_n^x,
	 \label{H_full}
\end{equation}
where $\sigma^z_n$ and $\sigma^x_n$ are the Pauli matrices acting on the $n$-th lattice site with $n = 1,...,N$ and
$N$ the total number of sites.
Here, $J$ denotes the spin-spin coupling and $h$ the transverse field.
For convenience we choose periodic boundary conditions: $\sigma_N^z \equiv \sigma_1^z$. This choice, however, does not influence our results.
Since discrete symmetries cannot be spontaneously broken at finite temperature in 1D systems characterized by short-range
interactions~\cite{landau2013course}, the model does not show a finite-temperature phase transition.
However, at zero temperature the system exhibits a quantum phase transition (QPT)~\cite{sachdev2007}
separating a ferromagnetic phase $(h<h_c)$ from a paramagnetic one $(h>h_c)$ where the critical point is given by $h_c=J/2$~\cite{pfeuty1970one}.

In the one-dimensional transverse-field Ising model with short-range interactions DQPTs have been studied already in great detail~\cite{pollmann2010,heyl2013,vajna2014,Abeling2016,ScalingHeyl,sharma2016slow,Puskarov2016,Heyl2017critical,dutta2018}.
It has been found that DQPTs occur whenever the system is quenched across the underlying equilibrium quantum critical point.
When interactions are long ranged, DQPTs can also occur but appear to be not any more linked to the model's quantum phase transition.
The DQPTs are rather either related to a different class of dynamical transitions in long-time steady states~\cite{zunkovic2016,jurcevic2016} or can occur even without known connection to a another class of (nonequilibrium) phase transitions~\cite{zauner2017probing,halimeh2017dynamical,homrighausen2017,lang2018}.
The latter class of DQPTs have acquired the notion of 'anomalous'.

As anticipated, in this work we consider the 1D transverse Ising chain subject to a quantum quench. Specifically, we prepare the system initially in the ground state
\begin{equation}
 | \psi \rangle = | \rightarrow \rangle := \bigotimes_{n=1}^N | \rightarrow \rangle_n \;\;\; | \rightarrow \rangle_n = \frac{1}{\sqrt{2}} [| \uparrow \rangle_n + | \downarrow\rangle_n],
 \label{eig_sigma_x}
\end{equation}
of our model at vanishing spin-spin coupling and nonzero external field corresponding to the initial Hamiltonian:
\begin{equation}
 H_i=H(t<0) = -h \sum_n \sigma_n^x.
\end{equation}
At time $t=0$, we switch to the opposite limit of $h=0$ and nonzero $J$ which yields:
\begin{equation}
 H_f=H(t>0) = -J \sum_n \sigma_n^z \sigma_{n+1}^z.
 \label{H_f}
\end{equation}
The quantum state evolves according to the final Hamiltonian, therefore at time $t$ it is given by:
\begin{equation}
 |\psi(t)\rangle = e^{itJ \sum_n \sigma_n^z \sigma_{n+1}^z} | \rightarrow \rangle.
\end{equation}
For such a quench, it is well known that the system exhibits DQPTs~\cite{heyl2013,vajna2014,ScalingHeyl}.
In Secs.~\ref{SBP} and~\ref{SPP} we also discuss deviations from this extreme protocol.

\section{Main Results} \label{sec:mainresults}
As the main result of this paper we introduce a dynamical analogue $f(m,t)$ of a free energy (section \ref{Dynamical Free Energy}) in the out of equilibrium
dynamics of the Ising model, with $m \in \mathbb{C}$ playing the role of the order parameter, which in our nonequilibrium setup becomes a complex magnetization.
While thermodynamic potentials in equilibrium statistical physics obey a minimizing principle, we find that the physically relevant region of
the effective free energy landscape $f(m,t)$ is determined by a saddle point instead:
\begin{equation}
	 f(t) = \underset{m\in\mathbb{C}}{\mathrm{sp}} f(m,t) \, ,
	 \label{f(t)}
\end{equation}
where $\mathrm{sp}$ stands for saddle point taken here over the set of complex magnetizations $m \in \mathbb{C}$. 
In the equilibrium limit where both $f(m,t)$ and $m$ are real, the saddle point turns back into the minimization principle, signaling a strong similarity of our construction to the equilibrium case.
We compute $f(m,t)$ via an equivalent effective potential $g(\mu,t)$.
\begin{equation}
	 f(m,t) = \underset{\mu\in\mathbb{C}}{\mathrm{sp}} [m\mu - g(\mu,t)] \, ,
	 \label{Legendre}
\end{equation}
which is the analog to Legendre transforming the thermodynamic potentials, with the magnetization $m$ and the field $\mu$ being conjugate variables.
The saddle point solution yields $\mu=\mu(m,t)$, which is nothing but the analog of an equation of state.

The DQPTs for the considered quantum quenches in the Ising model are known to be critical points associated with an unstable fixed point of an RG transformation and therefore to exhibit scaling and universality \cite{ScalingHeyl}.
This motivates us to study the behavior of $f(m,t)$ in the vicinity of the DQPT. 
We find that $f(m,t)$ admits an expansion in terms of $m$ yielding:
\begin{align}
  f(m,t) = & f_0(t) +\text{sign}(\theta) \alpha \tilde{m}^2(t) + \mathcal{O}\left( m^4(t) \right), \\ & \tilde{m}(t) = U(t)m,
  \label{fit_2}
\end{align}
where  $f_0(t)$ is a time-dependent function and $U(t)= \exp[i\theta(t)]$ represents a rotation in the complex plane. 
From the exact solution of the model we find that $\alpha$ is a time-independent constant and $\theta = (t_c - t)/t_c$ a linear function of the distance to the critical point in the vicinity of the DQPT at time $t_c$.

\section{DQPTs of the 1D Ising model} \label{DQPTs}

After having introduced our main results we now continue by discussing DQPTs for the considered nonequilibrium scenario and by outlining the main aspects of the methodology that is used in the remainder for the exact solution of the problem.

Applying the quantum quench protocol discussed in section \ref{model}, it is possible to write formally the Loschmidt amplitude, see Eq. \eqref{G}, as a conventional classical partition function \cite{ScalingHeyl}.
\begin{equation}
 \mathcal{G}(t) = \langle \rightarrow | e^{it\sum_n \sigma_n^z \sigma_{n+1}^z} | \rightarrow \rangle = \frac{1}{2^N}\text{Tr}\; e^\mathcal{H},
 \label{G_ising}
\end{equation}
where $\mathcal{H}$ is a complex Hamiltonian defined as:
\begin{equation}
 \mathcal{H} = K\sum_n \sigma_n^z \sigma_{n+1}^z, \quad K = itJ \, . 
 \label{H_nonhermitian}
\end{equation}
The only difference to the equilibrium case is that now the couplings can also be complex $K\in\mathbb{C}$.
This effective classical description offers various useful consequences such as exact solvability~\cite{sachdev2007} or the
construction of exact renormalization group transformations~\cite{nelson1975,huang2009}, as we will study in detail in the following.

\begin{figure}[tb!]
\includegraphics[width=0.9\columnwidth]{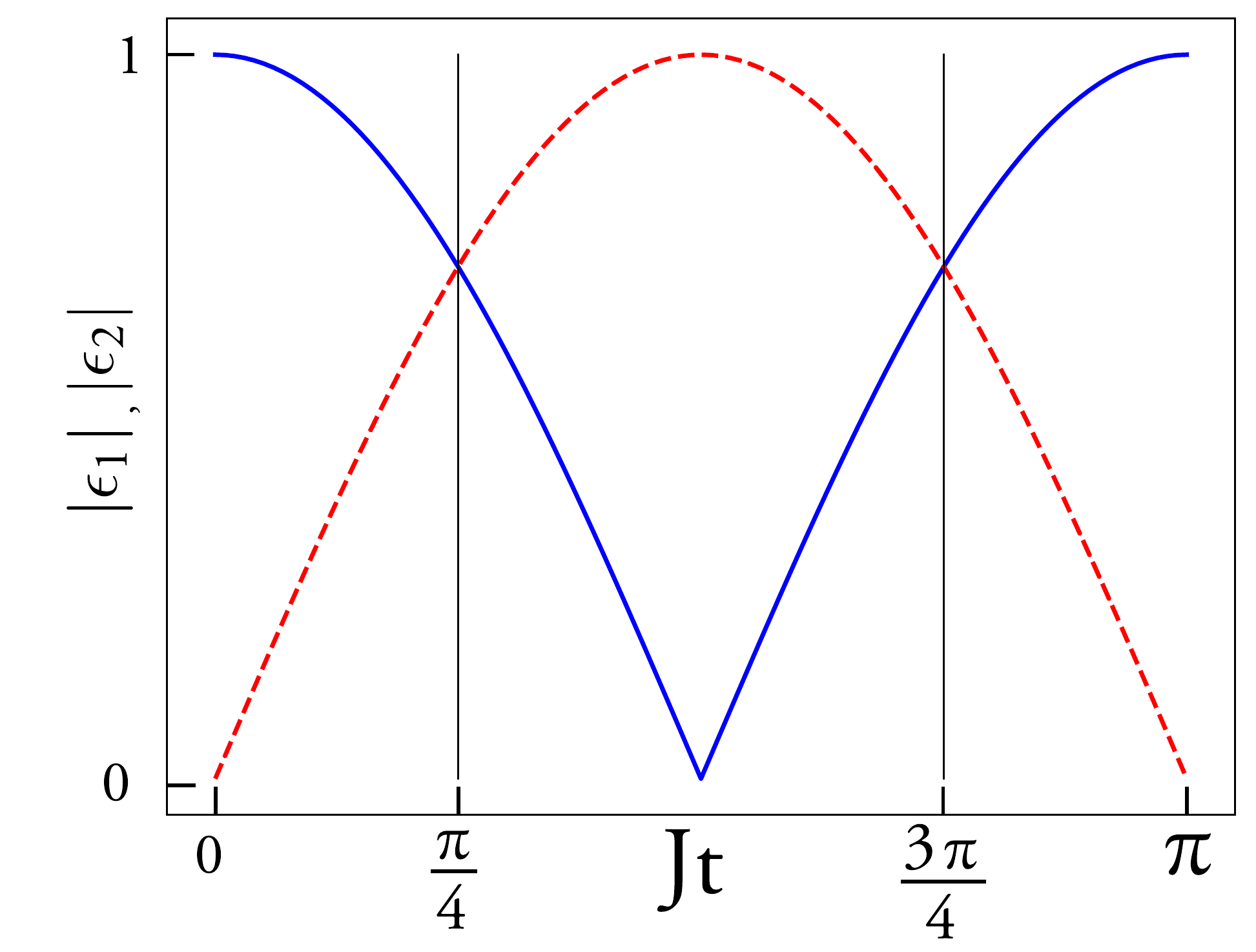}
\caption{Absolute values of the eigenvalues $\varepsilon_1, \varepsilon_2$ of the transfer matrix $T$ \eqref{D} for $\mu=0$. Vertical lines are at $t= \pi/4$ and $t=3\pi/4$ where the two eigenvalues cross each other and consequently the system undergoes a DQPT.}
\label{fig:eig_crossing}
\end{figure}

To see how it is possible to express formally the Loschmidt amplitude as a conventional classical partition function,
let us use the property that the initial state $|\psi\rangle = |\rightarrow \rangle$ is equivalent to an equally weighted linear combination of all spin configurations $|s\rangle = |s_1 \dots s_N\rangle$, $s_n=\uparrow \downarrow$ with $n=1,\dots,N$, along the ordering direction of the Ising model:
\begin{equation}
\begin{split}
  |\psi\rangle = \bigotimes_{n=1}^N  \left[ \frac{|\uparrow\rangle_n + |\downarrow\rangle_n}{\sqrt{2}} \right] = 
  2^{-\frac{N}{2}} \sum_s | s\rangle.
\end{split}  
\label{example_N}
\end{equation}
Since the operators in the Hamiltonian \eqref{H_nonhermitian} do not flip the z component of spins, the Loschmidt amplitude assumes a diagonal form in the z-spin configuration basis and therefore we recover Eq.~(\ref{G_ising}).
The partition function of the 1D Ising model is exactly solvable~\cite{huang2009,sachdev2007}, e.g. via
the transfer matrix technique~\cite{suzuki1985transfer}.
This procedure can be extended to complex couplings, providing a simple expression for the Loschmidt amplitude which
can be evaluated directly in the thermodynamic limit, i.e. $N\rightarrow \infty$.
For any $N$, the Loschmidt amplitude assumes the form:
\begin{equation}
 \mathcal{G} = \frac{1}{2^N} \text{Tr } T^N,
 \label{G_tech}
\end{equation}
where $T \in \mathbb{C} \times \mathbb{C}$ is a complex-valued transfer matrix. Before specifying the precise structure of $T$ let us slightly generalize our classical Hamiltonian $\mathcal{H}$ to
\begin{equation}
  \mathcal{H}_g = K \sum_n \sigma_n^z \sigma_{n+1}^z  + B \sum_n \sigma_n^z \, ,
  \label{H_classical} 
\end{equation}
which will turn out to appear naturally in various contexts discussed later on, e.g., when calculating the effective potential $g(\mu,t)$ appearing in Eq.~(\ref{Legendre}).
For this extended Hamiltonian the transfer matrix can be expressed as
\begin{equation}
\begin{aligned}
T = \left( \begin{array}{cc}
x^{-1}y^{-1} & x  \\
x & x^{-1}y  \end{array} \right),
\end{aligned}
\label{D}
\end{equation}
with
\begin{equation}
 \begin{split}
  x = e^{K},\quad  y = e^{B}.
  \label{xy}
 \end{split}
\end{equation}
As opposed to the equilibrium case with real-valued couplings, the transfer matrix $T$ is not hermitian in our case.
Nevertheless, $T$ can be expressed in terms of the right-left eigenvalues and eigenvectors, which allows to find 
an analytical expression for the Loschmidt amplitude \eqref{G_ising}:
\begin{equation}
 \mathcal{G}(t) = \frac{1}{2^N}  ( \varepsilon_1^N + \varepsilon_2^N),
 \label{G_ising_3}
\end{equation}
where $\varepsilon_1$ and $\varepsilon_2$ are the two right eigenvalues of $T$.
In the thermodynamic limit, the eigenvalue with largest magnitude is dominating:
\begin{equation}
 \mathcal{G}(t) = \lim_{N \rightarrow \infty} \frac{1}{2^N} \varepsilon^N ,
 \label{G_ising_4}
\end{equation}
where $\varepsilon = \varepsilon_1$ if $|\varepsilon_1| > |\varepsilon_2|$ or $\varepsilon = \varepsilon_2$ if $|\varepsilon_1| < |\varepsilon_2|$.
Thus in this model, nonanalytic structures occur at each time when there is a crossing between the absolute value of the eigenvalues $\varepsilon_1$ and $\varepsilon_2$
~\cite{andraschko2014}, see Fig. \ref{fig:eig_crossing}.

While the complex partition function and thus the Loschmidt amplitude can be computed exactly using the transfer matrix technique, one can obtain additional insights into the nature of the DQPTs by studying the behavior under a RG transformation.
This is particularly interesting in the limit $\mu=0$ where in this way it has been shown that the appearing DQPTs are critical points associated with an unstable fixed point~\cite{ScalingHeyl}.
By performing a decimation RG upon integrating out every second lattice site~\cite{nelson1975} one obtains the following recursion relation
\begin{equation}
  \tanh(K') = \tanh^2(K) \, ,
\end{equation}
determining the renormalized coupling $K'$ after one step in terms of the initial ones $K$.
This relation is valid both in and out of equilibrium with the only difference that in the dynamical context the coupling $K\in\mathbb{C}$.
This RG equation exhibits two fixed points $K^\ast = 0,\infty$ corresponding to the infinite and zero temperature fixed points of the equilibrium partition function.
When plugging in the critical value $K_c=\pi/4$ for the occurence of the DQPT, one directly finds that the coupling flows into the unstable $K^\ast=\infty$ fixed point implying scaling and universality~\cite{ScalingHeyl}.

\section{Effective Free Energy} \label{Dynamical Free Energy}

Before introducing the effective free energy for the DQPTs, it is useful to first outline how one can construct the free energy in equilibrium statistical physics for the Ising model.

\subsection{Equilibrium free energy of the Ising chain}
\label{sec:equilibriumFreeEnergy}

Let us express the partition function $Z(\beta)$ as:
\begin{align}\label{Z_T_classical}
	Z(\beta) & = e^{-\beta F(\beta)}= \mathrm{Tr}\; e^{ -\beta \mathcal{H}} = \\ \nonumber
	  & = \sum_s e^{ -\beta \mathcal{H}(s)} =  \int dM e^{-\beta F(M,\beta)}.
\end{align}
Here, $ \mathcal{H}(s)$ denotes the energy of a spin configuration $ s =(s_1,s_2,...,s_N)$ with $s_n = \pm 1$ and 
\begin{equation}
	e^{-\beta F(M,\beta)} = \sum_s e^{-\beta \mathcal{H}(s)} \delta \left(M - \sum_n s_n \right),
	\label{exp_free_en}
\end{equation}
defines the free energy $F(M,\beta)$ at a fixed magnetization $M$.
Both $F(M,\beta)$ and $M$ are extensive, hence it is possible to express them  in terms of intensive quantities: $F(M,\beta) = Nf(m,\beta)$, $m=M/N$ where $N$ is the number of lattice sites.

In the thermodynamic limit, the free energy density $ f(\beta)= F(\beta)/N$  in Eq. \eqref{Z_T_classical}, is given by:
\begin{equation}
	f(\beta)= \min_{m} f(m,\beta) = f(m^*(\beta),\beta),
	\label{f}
\end{equation}
recovering the minimization principle of the free energy by applying the Laplace's method to the integral in Eq. \eqref{Z_T_classical}. Here, $m^*(\beta)$ denotes the magnetization that fulfills the minimization condition.
 
For later convenience, it will turn out to be useful to consider also the Laplace transform of $e^{-Nf(m,\beta)}$ denoted by $G(\mu,\beta)$:
\begin{equation}
G(\mu,\beta) = \int dm \, e^{N\mu m} e^{-Nf(m,\beta)}.
\label{G_mu}
\end{equation}
Using general results from large-deviation theory~\cite{touchette2009large} the function $G(\mu,\beta)$, analogous to the Loschmidt amplitude $\mathcal{G}(t)$, exhibits a particular functional dependence on system size $N$:
\begin{equation}
G(\mu,\beta) = e^{Ng(\mu,\beta)} \, ,
\end{equation}
with $g(\mu,\beta)$ an intensive function.
By applying again Laplace's method to the integral in Eq. \eqref{G_mu}, the relation between $g(\mu,\beta)$ and $f(m,\beta)$ is, up to corrections vanishing for $N\rightarrow \infty$, a Legendre transformation~\cite{touchette2009large}:
\begin{equation}
	g(\mu,\beta) = \sup_{m}  \left[ \mu m-f(m,\beta) \right] .
	\label{LT_g}
\end{equation}
Equivalently, we get that:
\begin{equation}
	f(m,\beta) = \sup_{\mu}  \left[ \mu m  - g(\mu,\beta) \right] ,
	\label{LT}
\end{equation}
since the inverse of the Legendre transform is a Legendre transform itself.
Solving Eq. \eqref{LT_g} for its supremum, i.e., $\mu = df(m,\mu)/dm$, gives $m=m(\mu,\beta)$ which is nothing but the equation of state (EoS) relating the two conjugate quantities  magnetization density $m$ and  field $\mu$.

\subsection{Effective free energies for DQPTs}

Now, let us extend these concepts to the dynamical analogue of the partition function in Eq. \eqref{G_ising}. We know that we can formally consider the Loschmidt amplitude as a
 complex partition function, see Eq. \eqref{G_tech}, meaning that the formalism introduced in Section \ref{sec:equilibriumFreeEnergy} can be used also in the nonequilibrium context, taking into account that the Hamiltonian $ \mathcal{H}$ is now complex, see Eq. \eqref{H_nonhermitian}.
Using the definition of $G(\mu,\beta)$ in Eq. \eqref{G_mu} and Eq. \eqref{exp_free_en}, we find that we can express $G(\mu,t)$ in the following way:
\begin{equation}
	G(\mu,t) = \sum_s e^{ \mathcal{H}(s,\mu,t)}, \;\;\; \mathcal{H}(s,\mu,t) = \mathcal{H}(s,t) -  \mu \sum_n s_n.
	\label{G_complex}
\end{equation}
Therefore, $G(\mu,t)$ is again a classical partition function with $\mathcal{H}(s,\mu,t)$ a classical 1D Ising Hamiltonian with complex parameters. $G(\mu,t)$ can be exactly solved 
using the method introduced in Section  \ref{DQPTs} , see Eqs.~\eqref{G_tech},\eqref{G_ising_3},\eqref{G_ising_4}, leading in the thermodynamic limit $N \rightarrow \infty$ to the expression:
\begin{equation}
 G(\mu,t) = \varepsilon_1^N + \varepsilon_2^N \underset{N \rightarrow \infty}{\overset{}{\longrightarrow}}   \varepsilon^N \, ,
 \label{G(mu)}
\end{equation}
with $\varepsilon_{1/2}$ the two right eigenvalues of the transfermatrix with $\varepsilon = \varepsilon_1$ when $|\varepsilon_1|>|\varepsilon_2|$ or $\varepsilon = \varepsilon_2$ otherwise.
Consequently, $g(\mu,t)$ is determined solely by the right eigenvalues of the transfer matrix:
\begin{equation}
	g(\mu,t) = \frac{1}{N}\log\;G(\mu,t) \asymp \frac{1}{N}\log(\varepsilon^N) = \log(\varepsilon).
	\label{g_f}
\end{equation}
It remains now to compute the actually targeted quantity $f(m,t)$.
Suppose we would already know $f(m,t)$, then we can use a saddle-point approximation in the thermodynamic limit $N\to\infty$ to determine $g(\mu,t)$ according to Eq.~(\ref{G_mu})
\begin{equation}
    g(\mu,t) = \underset{m\in\mathbb{C}}{\mathrm{sp}}[\mu m-f(m,t)].
\end{equation}
The saddle-point condition can be equivalently formulated as
\begin{equation}
    \mu = \frac{df(m,t)}{dm} \Rightarrow m = m(\mu,t).
\end{equation}
which is the out-of-equilibrium counterpart to the equation of state relating the two conjugate quantities $m$ and $\mu$.
These formulas, similar to Legendre transforms, can be inverted to yield
\begin{equation}
  f(m,t) = \underset{\mu\in\mathbb{C}}{\mathrm{sp}}[\mu m-g(\mu,t)].
  \label{eq_fm_sp}
\end{equation}
The resulting equation of state has a particularly physically transparent form:
\begin{equation}
m = \frac{dg(\mu,t)}{d\mu} =  -\frac{1}{N}\sum_n \langle s_n^z\rangle_{\mu},
\label{SPA_T}
\end{equation}
with
\begin{equation}
    \langle s_n^z\rangle_{\mu}= \frac{1}{G(\mu,t)} \mathrm{Tr} \left[ s^z_n \,\,  e^{\mathcal{H}(s,\mu,t)}\right],
    \label{SPA_T_2}
\end{equation}
the local magnetization of a classical Ising model in the presence of a (complex) longitudinal field $\mu$ and $G(\mu,t)$ the analogue to the partition function.
Transfer matrix techniques allow to compute Eq. \eqref {SPA_T}  exactly~\cite{sachdev2007} at any $N$.
Details on the calculations are provided in Appendix~\ref{Appendix}.
Using then Eq.~(\ref{eq_fm_sp}) we can readily obtain the desired effective free energy density $f(m,t)$.

In general we find that at a fixed complex magnetization density $m$ there are multiple values of $\mu$ satisfying Eq. \eqref {SPA_T}, implying that there exist several saddle points located at different complex fields $\mu_i$, say.
In the calculation of the effective free energy we keep the value $\mu_\ast$ of all $\mu_i$ which contributes dominantly.
Specifically, in case of multiple saddle points we have
\begin{equation}
  e^{-Nf(m,t)} = \sum_i e^{-N [ \mu_i m-g( \mu_i,t)] } \stackrel{N\to\infty}{\longrightarrow} e^{-N[ \mu_\ast m-g( \mu_\ast) ]} \, ,
\end{equation}
where $\mu_\ast$ is chosen such that $ \mathrm{Re} [\mu_\ast m-g( \mu_\ast)  ] < \mathrm{Re} [\mu_i m-g( \mu_i) ]$ for all $i\not= \ast$.
In this case all the other contributions are exponentially suppressed in system size.

\begin{figure}[tb!]
	\includegraphics[width=1.02\columnwidth]{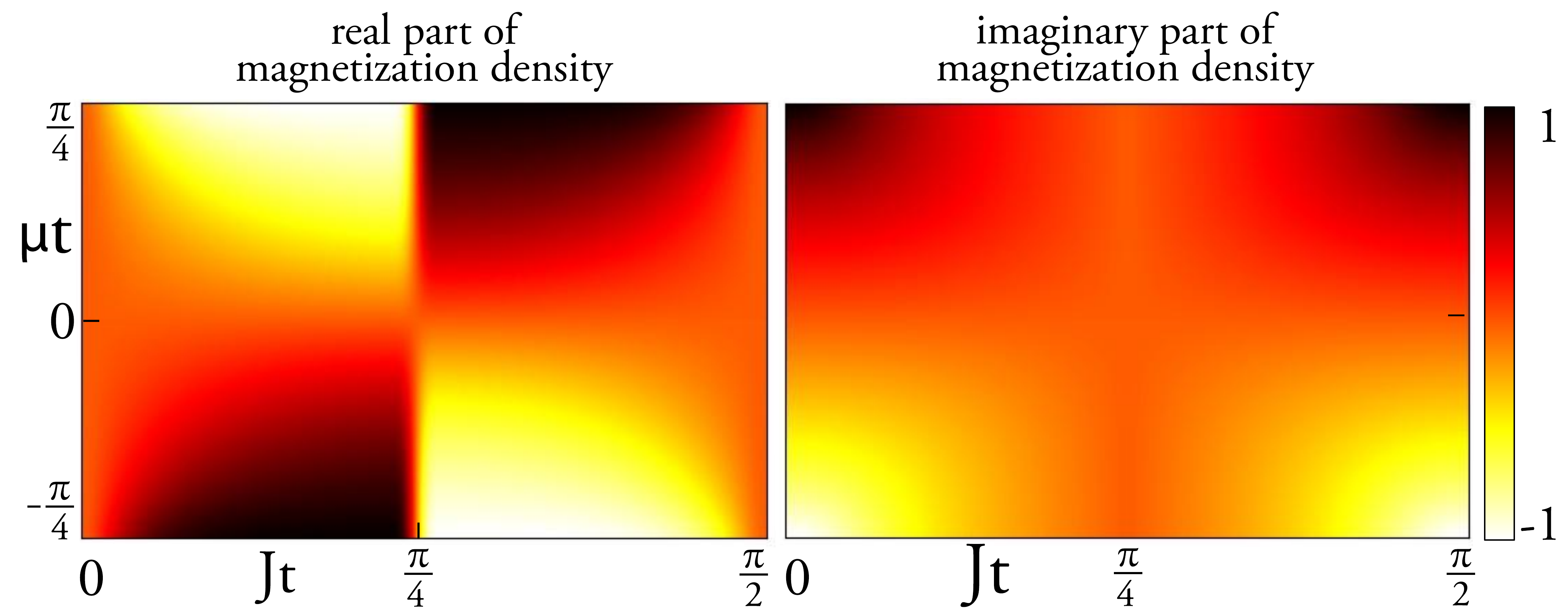}
	\caption{Dynamical analog of the equation of state for the complex magnetization density $m(\mu,t)$ as a function of the complex magnetic field $\mu$ for different times $t$. Left: Real part of the magnetization density as a function of $Jt$ and $\mu t$. Crossing the vertical line $Jt=\pi/4$ the magnetization	density has a jump if the external field $\mu$ is different from zero. This reflects the nature of the DQPT occuring at $Jt=\pi/4$, which is continuous-like at vanishing field $\mu$ and first order otherwise. The sudden jump comes from a switching of the saddle point in Eq.~\eqref{G(mu)}, which affects $m$ via Eqs.~\eqref{SPA_T},~\eqref{SPA_T_2}.
		 Right: Imaginary part of the magnetization density as a function of $Jt$ and $\mu t$. This
		quantity does not have any nonanalyticities in the $Jt$-$\mu t$ plane.}
	\label{fig:magnetization_diagram_real}
\end{figure}

In Fig.~\ref{fig:magnetization_diagram_real} we show data for the equation of state $m(\mu,t)$ for the case of a real-valued field $\mu\in\mathbb{R}$.
As one can clearly see, there appears a sharp structure at $Jt=\pi/4$, where for $\mu=0$ the system experiences a continuous DQPT~\cite{ScalingHeyl}.
At nonzero $\mu\not=0$ we observe that the real part of $m(\mu,t)$ develops a discontinuous behavior which we discuss in more detail in Sec.~\ref{SBP}.
The imaginary part $m_I(\mu,t)=\mathrm{Im}[ m(\mu,t)]$ on the other hand appears to be continuous as a function of time across $Jt=\pi/4$.
The observed sharp change in $m(\mu,t)$ originates in a switching of the dominant saddle point.
Accordinly, this directly translates into a sharp changes in the effective free energy density $f(m,t)$ which we show in Fig. \ref{fig:NO_jumps_real_free}.
There, we plot our final result for the real part $f_R(m,t)$ of the effective free energy density ($f(m,t)=f_R(m,t)+ if_I(m,t)$) for two different times in the vicinity of the critical time $Jt_c = \pi/4$ of the DQPTs, one for a time $t<t_c$ shortly before the DQPT and for $t>t_c$.
As one can observe clearly, the DQPT reflects in as a sudden $\pi/2$ rotation across $t_c$.
While we have illustrated the equation of state in Fig.~\ref{fig:magnetization_diagram_real} for the case of a real field $\mu$, let us note that the full effective free energy density $f(m,t)$ is obtained from the dominant saddle point in the whole complex $\mu \in \mathbb{C}$ plane.

Finally, let us note how the full analog of a free energy $f(t)$ can be obtained from the knowledge of $f(m,t)$.
Following Eq.~\eqref{Z_T_classical} we again observe that in the out-of-equilibrium context the integral over the magnetization in the thermodynamic limit is dominated by a saddle point:
\begin{equation}
  f(t) =\underset{m\in\mathbb{C}}{\mathrm{sp}} f(m,t) \, ,
  \label{f_saddle_point}
\end{equation}
which is the formula outlined already in Sec.~\ref{sec:mainresults}.
Again we find that in general there exists not a single saddle point, so that we always take only the dominant contribution analogous to before. 
As one can see from our data in Fig.~\ref{fig:NO_jumps_real_free} one can directly identify a clear saddle point around $m=0$, which we is central for expanding $f(m)$ in powers of $m$ as we discuss in the following section.

\section{Expanding the effective free energy} \label{Landau} \label{Magnetization density}
On the basis of the calculations in the previous section it is the aim of the following to study expansions of the effective free energy in powers of the complex magnetization~\eqref{SPA_T_2}, which plays the role of the order parameter in analogy to conventional Landau theory.
We have already seen that DQPTs show typical features of equilibrium phase transitions, such as the nonanalytic behavior of the effective free energy $f(t)$ at the critical time.
What motivates the targeted expansion in particular, is that it is known from an exact RG analysis that the studied DQPTs are of continuous nature~\cite{ScalingHeyl} which makes an analog to Landau theory particularly promising.

\begin{figure}[tb!]
	\includegraphics[width=1.02\columnwidth]{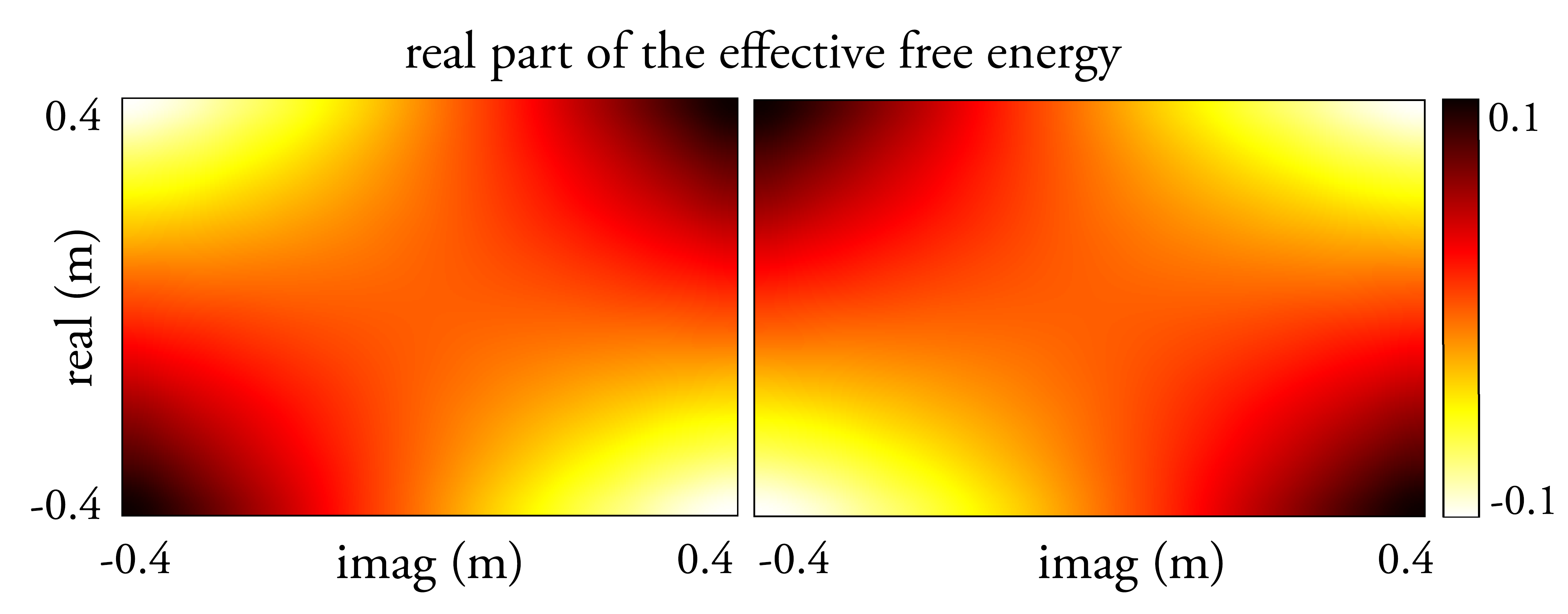} \hfil
	\caption{Real part of the effective free energy at different times in the complex magnetization density plane. Within the range of the displayed $m\in\mathbb{C}$ the effective free energy $f(m,t)$ does not exhibit sudden jumps as happen in Fig.~\ref{fig:jumps_free_en}. This is a consequence of the magnetization range which  is chosen in such a way that $g(\mu,t)$ in Eq.~\eqref{g_f} and therefore $f(m,t)$ is obtained through either $\varepsilon_1$ or $\varepsilon_2$ which are continuous functions of their variables. 
	The plot on the left is obtained with $t<t_c$, in particular $t =\frac{\pi}{4}-\frac{\pi}{32}$, while the plot on  the right with $t>t_c$, in particular $t =\frac{\pi}{4}+\frac{\pi}{32}$.}
	\label{fig:NO_jumps_real_free}
\end{figure}

Let us again discuss Fig.~\ref{fig:NO_jumps_real_free} which shows the real part $f_R(m,t)$ of the effective free energy density $f(m,t)$ as a function of the complex magnetization density $m$ at
two different times $t$ close to $t_c=\frac{\pi}{4}$ of the DQPT: one value $t=\frac{\pi}{4}-\frac{\pi}{32}<t_c$ and one at $t=\frac{\pi}{4}+\frac{\pi}{32}>t_c$.
We observe that for $t<t_c$ the minima are in the corners of the first and third quadrant, while the maxima in the other two corners.
As the time increases, the positions of the maxima and minima rotate counterclockwise (not shown), but at time $t=t_c$, there is an additional sudden rotation of $\pi/2$.
This rotation one can clearly indentify in Fig.~\ref{fig:NO_jumps_real_free}.
Analogous considerations hold for the imaginary part.

Throughout all times we observe a saddle point at $m=0$ suggesting that an expansion in powers of $m$ is suitable.
We find that $f(m,t)$ admits the following expansion:
\begin{equation}
\begin{aligned}
f(m,t)  = f_0(t) + \text{sign}(\theta) \alpha \tilde{m}^2(t)  + \mathcal{O}(m^4), 
\end{aligned}
\label{fit_f_m}
\end{equation}
with
\begin{equation}
\tilde{m} = U(t) m, \quad U(t) = e^{i\theta(t)}\, .
\end{equation}
Here, $f_0(t)$ is a time-dependent function and $\alpha$ is a constant in the vicinity of the DQPT at time $t_c$ whose value can be determined from Eq. \eqref{Legendre} with Eq. \eqref{fit_f_m}.
The time-dependence of the angle $\theta(t)$ incorporates the slow counter clockwise rotation mentioned already before.
In the vicinity of the DQPT we extract from the numerical solution a linear dependence of the angle:
\begin{equation}
\theta(t) \stackrel{t\to t_c}{\longrightarrow} \frac{t_c - t}{t_c} \; .
\label{theta}
\end{equation}
The sudden rotation of $\pi/2$ observed at the critical time $t_c$ is not caused by the slow rotation $e^{i\theta(t)}$ but rather by the change of sign of $\theta$
 encoded in sign($\theta)$, which switch the maxima of the effective free energy with the minima and vice versa.
In particular, we find that this sudden change originates from a switching of the dominant saddle point.

Now let us discuss the obtained results in light of conventional Landau theory.
At least in high dimensions one might expect competing quadratic and quartic powers of the order parameter appearing in the expansion of the free energy which upon crossing a phase transition transforms the free energy from having a single minimum to a double well landscape.
In the present example, however, the quartic term turns out to not contribute significantly ruling out a competition as one might expect from the conventional picture.
On a formal level, this can be attributed to our observation that $\alpha$ is to leading order in the temporal distance to the critical point a nonvanishing constant which leaves the quartic magnetization density contribution always subleading.
However, let us note that this might also be attributed to the low dimensionality of the system.
The classical 1D Ising model does not exhibit an extended symmetry-broken phase, but rather only a singular point showing ferromagnetic order, namely at temperature $T=0$.
There, the ground state is doubly degenerate and the free energy $f(m)$ is finite only on the two points $m=\pm 1/2$ and infinite otherwise.
In this sense it exhibits an analog to a double well free energy landscape.
At any nonzero deviation from this singular $T=0$ point, however, the free energy exhibits only a single minimum at $m=0$, since the system immediately enters a disordered phase.
Our observations for the dynamical case are completely analogous to the equilibrium phenomenology in that for any nonzero deviation from the critical point the free energy is dominated by the quadratic contribution.
What is different, however, is, that $\alpha$ nevertheless vanishes in the $T\to0$ limit in equilibrium, which we don't observe in the dynamical context studied here.
We conclude that a nontrivial expansion of $f(m,t)$ might require at least two dimensions where the symmetry-broken phase is extended and not just a singular point.

\section{Influence of a symmetry-breaking perturbation} \label{SBP}
So far we have studied the properties of the system in a special case for a quantum quench in the transverse-field Ising chain for vanishing initial coupling and vanishing final transverse field.
 Now we aim to study the influence of a longitudinal field in the final Hamiltonian $H$, which in equilibrium is a relevant perturbation, onto the nature of the DQPTs.
Let us therefore suppose that the final Hamiltonian $H_f$ is given by:
\begin{equation}
\begin{split}
H_f =- J\sum_n \sigma_n^z \sigma_{n+1}^z -\mu \sum_n \sigma_n^z.
\end{split}
\label{H_tilde}
\end{equation}
From the construction in Sec.~\ref{DQPTs} it follows immediately that the Loschmidt amplitude can still be mapped onto classical partition function given by:
\begin{equation}
\mathcal{G}(t) = \mathrm{Tr} e^{\mathcal{H}} \, , \quad \mathcal{H} = K\sum_n \sigma_n^z \sigma_{n+1}^z + B \sum_n \sigma_n^z \, ,
\end{equation}
with $K=iJt$ and $B=i \mu t$.

\begin{figure}[tb!]
	\includegraphics[width=0.95\columnwidth]{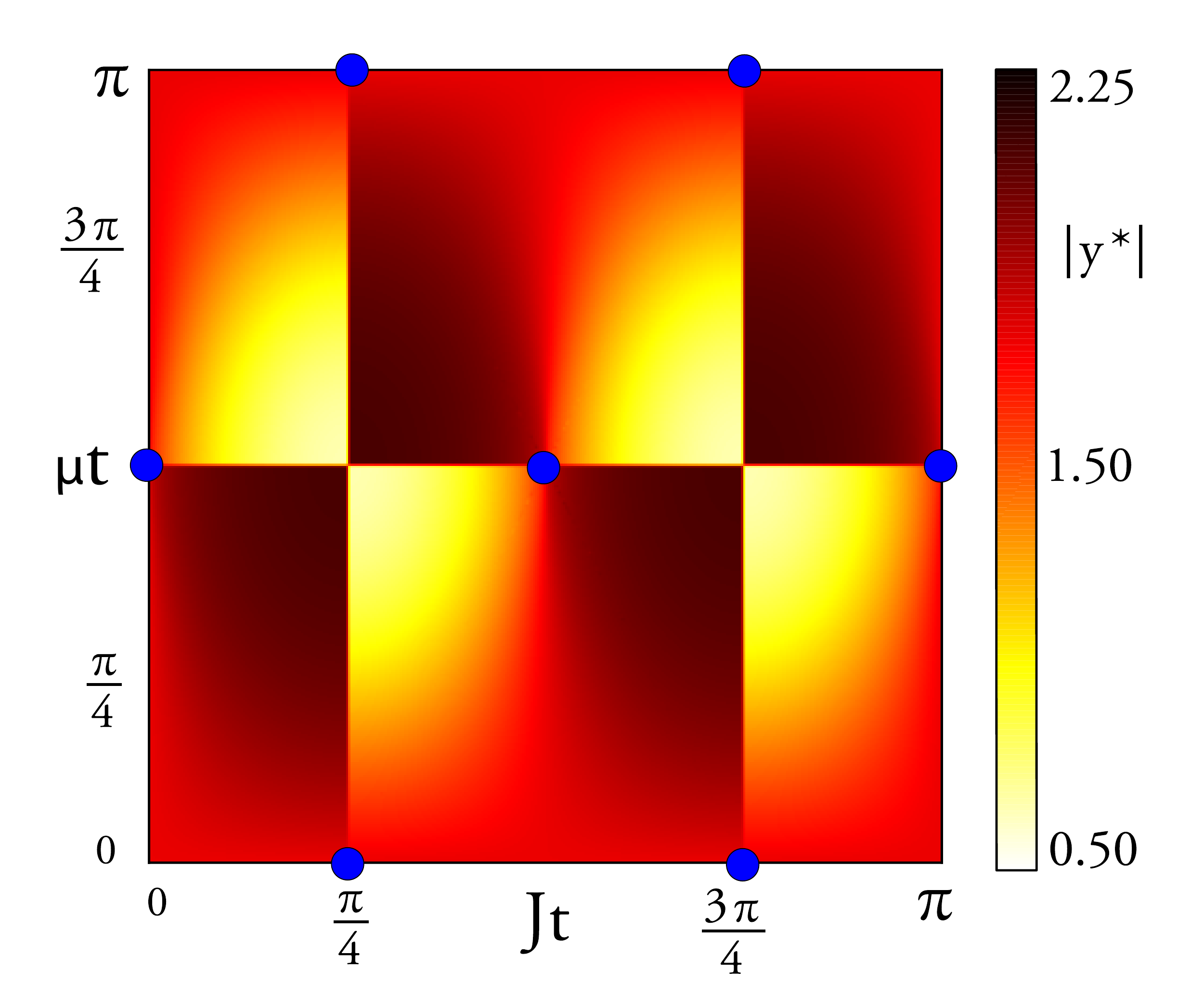}
	\caption{Absolute value of the fixed point of the $y$ variable. The blue points indicate the couple 
		of couplings ($\mu t,Jt$) which flow to the unstable fixed point.}
	\label{fig:RG_plane_x_real}
\end{figure}

In the following we now aim to study the fixed points of this resulting partition function using a standard decimation RGs~\cite{stanley1999scaling,nelson1975,huang2009}, which can be equally used for the complex couplings appearing in our case.
A suitable choice for the RG parametrization is analogous to the one used in  Eq. \eqref{xy} with $x=e^K$ and $y=e^B$.
Performing one RG step the system is described by renormalized couplings and fields and therefore new variables $ x' = e^{K'},y' = e^{B'} $ which are related to $x,y$ by the RG equations~\cite{huang2009,nelson1975}:
\begin{equation}
\begin{split}
x' = \frac{(y+1/y)^{1/2}}{(x^4+1/x^4+y^2+1/y^2)^{1/4} } \\
y' = \frac{(x^4+y^2)^{1/2}}{(x^4+1/y^2)^{1/2}}.
\end{split}
\label{RG}
\end{equation}
In Fig.~\ref{fig:RG_plane_x_real} we show the numerically obtained value $|y^\ast|$ when the RG has converged for varying initial conditions, which captures our main findings that we aim to explain in the following.

The above set of coupled RG equations exhibits two different classes of fixed points: 
\begin{equation}
(x^* = 1; \forall y^*) \; \longrightarrow \text{stable fixed point},
\label{stable}
\end{equation}
\begin{equation}
(x^* = 0; y^*=1) \; \longrightarrow \text{unstable fixed point}.
\label{unstable}
\end{equation}
This is the same set of fixed points one encounters also in the equilibrium case when starting from purely real couplings. 
There is a line of stable fixed points corresponding to vanishing spin-spin coupling $K=0$ for any value $\B$ for the field.
On this line of fixed points the physics is equivalent to uncorrelated spins in an external field implying a vanishing correlation length.
On the contrary, the unstable fixed point is characterized by a divergent spin coupling $K\to\infty$ at vanishing field $B=0$, or in a more general sense $B=2\pi n i$ with $n\in\mathbb{Z}$.
This fixed point corresponds to the zero-temperature limit of the Ising model without external field, which exhibits order in terms of a nonzero magnetization.

\begin{figure}[tb!]
	\includegraphics[width=0.99\columnwidth]{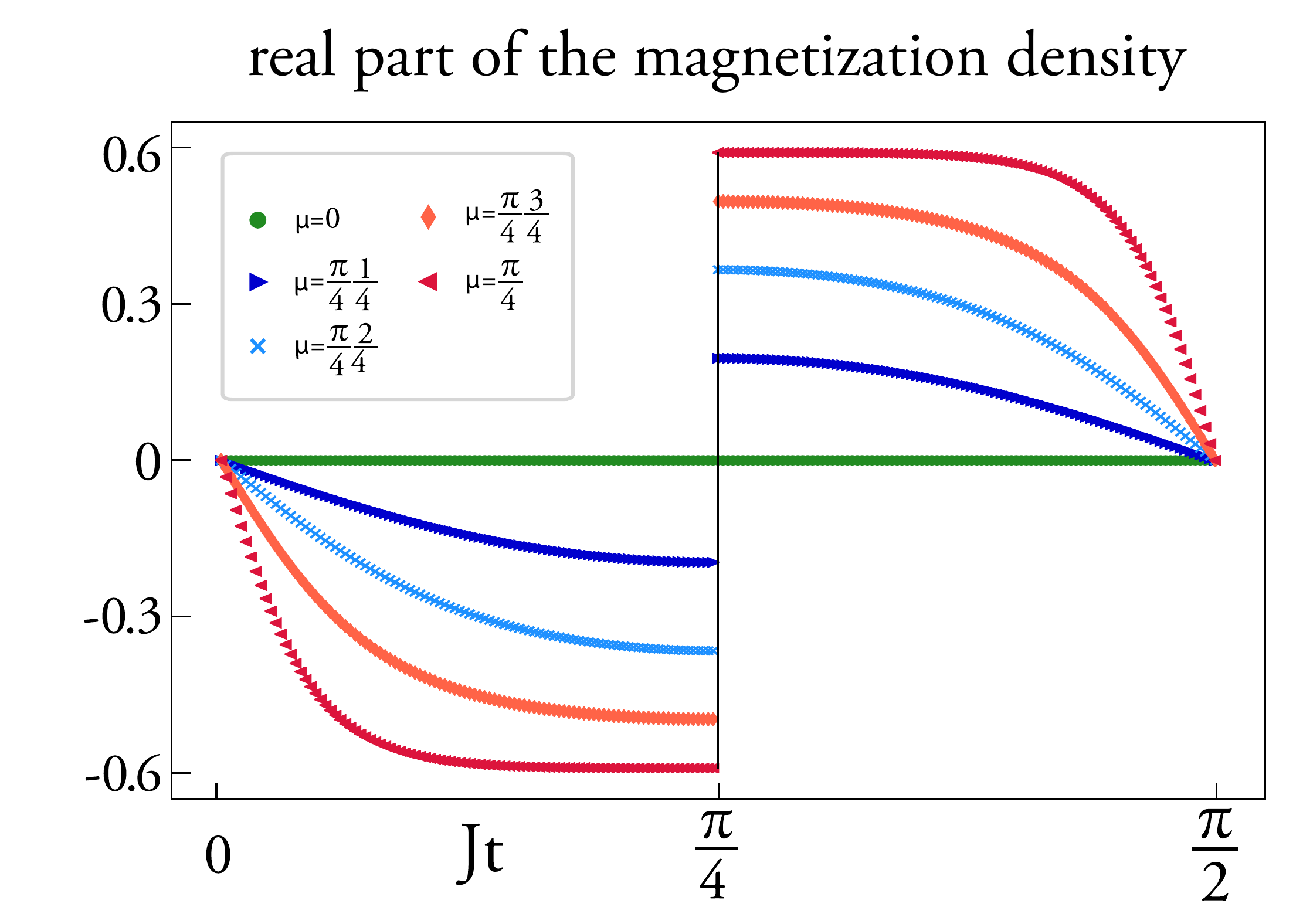} \hfil
	\caption{ Real part of the magnetization density $m(t)$ obtained from the EoS~\eqref{SPA_T} as a function of $Jt$ for different values of $\mu$. The jump of $m(t)$ at the critical time $Jt=\pi/4$ decreases linearly with the field $\mu$. This feature outlines the nature of the DQPT which is first-order like at nonzero field $\mu$, while it is continuous at vanishing field $\mu=0$.}
	\label{fig:magnetization_scatter}
\end{figure}

Depending on the initial conditions $(K,B)$ for the RG transformation, the system ends up in different fixed points, see Fig.~\ref{fig:RG_plane_x_real}.
Because the Pauli matrices that appear in the Hamiltonian have eigenvalues $\pm 1$ and because of the exponential structure of the Boltzmann weights, the complex partition function is symmetric under $\mu \mapsto \mu+n\pi$ and $J\mapsto J+n\pi$ for any $n\in\mathbb{Z}$.
Along the line of vanishing field $B=0$ we recover the DQPT at $Jt=\pi/4$ or $Jt=3\pi/4$ associated with the unstable fixed point of the unperturbed Ising model studied before, which is indicated in the figure with a blue dot.
As it happens in equilibrium~\cite{nelson1975}, a nonvanishing  field $B\not=0$ in general is a relevant perturbation attracting the system to a different stable fixed point.
However, we find in agreement with previous work on quantum quenches that a DQPT still exists~\cite{karrasch2013dynamical}.
As one can clearly see by following the vertical line along $Jt=\pi/4$ or $Jt=3\pi/4$ as a function of the field $\mu$, we find that $|y^\ast|$ is not anymore continuous but rather acquires a jump upon crossing the $Jt=\pi/4$ line.
In Fig.~\ref{fig:magnetization_scatter} we show data for individual cuts for fixed magnetic fields $\mu$ showing again the jumps in the magnetization.
As a consequence the DQPT turns from second to first order.
Remarkably, however, there appear still specific isolated points in the $J$-$\mu$ plane where the DQPT becomes again continuous although the field is nonzero as opposed to the equilibrium case where this is not possible.
These specific points are located at $\mu t=\pi/2$ when $Jt=0,\pi/2$, as indicated by the blue dots in Fig.~\ref{fig:RG_plane_x_real}.

As a consequence of the explicitly broken symmetry the saddle point of the effective free energy $f(m)$ shifts to a nonzero value $m\not=0$ implying that the expansion in the spirit of a Landau theory acquires a linear in $m$ contribution.
Surprisingly, it can, however, happen for larger fine-tuned fields that one of the specific continuous DQPTs for $\mu t=\pi/2$ are again encountered rendering the effective free energy analogous to the continuous DQPTs studied already before.

\section{Influence of symmetry-preserving perturbations} \label{SPP}
The perturbations considered in the previous section break the symmetry of the Hamiltonian, while 
we now aim to explore what happens when adding symmetry-preserving perturbations.
For that purpose we use a perturbative analytical approach to map the Loschmidt amplitude again onto classical partition functions, which, however, might contain additional couplings compared to the unperturbed one.
Using RG transformations we then study the influence of these perturbations, in particular, whether they turn out to be relevant or not.

We start with perturbations that preserve on a Hamiltonian level the $\mathbb{Z}_2$ symmetry of the model.
Concretely, we consider in the following a Hamiltonian \eqref{H_full} with a weak nonvanishing initial coupling $ J_0 \ll h_0$ and a final Hamiltonian containing also a weak transverse field $h \ll J$.

\subsection{Perturbations of the initial condition}
Let us first discuss the influence of the perturbed initial condition.
The initial state, which is the ground state of Eq. \eqref{H_full} before the quench, we approximate using an adapted Schrieffer-Wolff (SW) transformation.
Consider Hamiltonian $H$ given as a sum of an unperturbed one, $H_0$, and a perturbation $\lambda V$ with $\lambda$ weak:
\begin{equation}
H = H_0 + \lambda V.
\end{equation}
Since here we are interested in how the ground state is modified after the introduction of the perturbation, we aim to find a unitary transformation $U=e^S$ analogous to a Schrieffer-Wolff transformation~\cite{bravyi2011schrieffer} such that the eigenvalue equation for the ground state: $ H |\psi' \rangle = E_0 |\psi' \rangle$ is transformed into $U H U^\dagger |\psi\rangle = E_0 |\psi\rangle$ with $|\psi \rangle = U|\psi'\rangle$ and $ UHU^\dagger = H _0+ \mathcal{O}(\lambda^2)$ within a perturbative construction.
In other words, we only require to find a transformed ground state and we do not aim at finding the full diagonalized Hamiltonian.
In the spirit of a SW transformation, this can be achieved by $U=e^S$ with:
\begin{equation}
 [S,H_0] | \psi \rangle =  \lambda V | \psi \rangle.
\label{S-W_GS}
\end{equation}
To corrections of the order $\mathcal{O}(\lambda^2)$, the eigenstate $|\psi \rangle $ of $H$ can be obtained via:
\begin{equation}
|\psi' \rangle = e^{-S}| \psi \rangle,
\end{equation}
which in the limit $J_0 \ll h_0$ can be solved to yield:
\begin{equation}
 |\psi'\rangle = \exp \left(\frac{iJ_0}{4h_0}\sum_n \sigma_n^z \sigma_{n+1}^z \right) |\psi\rangle.
 \label{Schrieffer}
\end{equation}
As a consequence of the introduction of a small spin-spin coupling in the Hamiltonian $H_0$, the new ground state is obtained flipping nearest neighbors spins of the unperturbed fully polarized ground state $|\psi \rangle$.
For the Loschmidt echo $\mathcal{G}(t)$ the result in Eq.~(\ref{Schrieffer}) implies that $\mathcal{G}(t) = \langle \psi' | e^{-iHt} | \psi'\rangle =  \langle \psi | e^{-iHt} | \psi\rangle$ because $[S,H]=0$.
Thus, perturbations to the initial condition don't contribute in lowest order perturbation theory, but rather only enter when taking higher order corrections into account.
Although straightforwardly possible to incorporate, it is beyond of the scope of the lowest-order perturbation theory we address here.

\subsection{Perturbations to the final Hamiltonian}\label{perturbations_final}

After having discussed how to deal with perturbed initial states, we now turn to perturbations in the final Hamiltonian.
In the limit where $h \ll J$, the Loschmidt amplitude \eqref{G}, which contains the 1D nearest neighbor Ising quantum Hamiltonian \eqref{H_full}, can be approximated to first order in $h/J$ as
a formal partition function described by a classical Ising Hamiltonian with next to nearest neighbor interactions using cumulant expansion methods~\cite{ScalingHeyl,schmitt2017quantum}. 
To obtain this result, we again decompose on a general level our Hamiltonian $H=H_0+\lambda V$ with $H_0 = -J\sum_n \sigma_n^z\sigma_{n+1}^z$ and $\lambda V = -h\sum_n \sigma_n^x$. It is then convenient to move from the Schr\"odinger  to the interaction picture, where the time-propagator can be written as:
$ e^{-iHt}=e^{-iH_0t}W(t)$ with $W(t) = \mathcal{T}\exp\left( -i \int_0^t dt' \lambda V(t')\right)$ and $V(t) = e^{iH_0t}V e^{-iH_0t}$ such that we get for the time evolved state:
\begin{align}
		|\psi(t)\rangle = & \exp(-i H t) |\psi\rangle = \exp(-i H_0 t)W(t) |\psi\rangle = \\
		 = & \sum_s e^{-iE_st}|s \rangle \langle s| \mathcal{T} e^{-i \int_0^t dt' \lambda V(t')} |\psi\rangle \, ,
	\label{psi_cumulant}
\end{align}
where $|s\rangle$ denotes the set of all spin configurations which are the eigenstates of $H_0$ with eigenenergies $E_0(s)$. 
Using this equation we can now perform a cumulant expansion which up to lowest order in $\lambda$ yields~\cite{schmitt2017quantum}:
\begin{equation}
	\frac{\langle s| \mathcal{T} e^{-i \int_0^t dt' \lambda V(t')}| \psi\rangle}{\langle s| \psi\rangle}  = e^{-i\int_0^t dt' \, \frac{\langle s| \lambda V(t')\psi\rangle}{\langle s| \psi\rangle} + \mathcal{O}(\lambda^2) }.
	\label{Cumulant}
\end{equation}
Defining $\overline{V}(s,t) =i \int_0^t dt' \, \langle s| \lambda V(t')\psi\rangle / \langle s| \psi\rangle $  we obtain the following form for the Loschmidt amplitude for the fully polarized initial condition:
\begin{equation}
\begin{split}
\mathcal{G}(t) =\frac{1}{2^N} \sum_s e^{-itE_0(s)-\overline{V}(s,t)}= \frac{1}{2^N}{\rm Tr} \; e^{\tilde{H}},\\
\end{split}
\label{H_NNN}
\end{equation}
with $\tilde{H}(s,t) = -itE_0(s)-\overline{V}(s,t)$.
For the concrete case of an Ising chain in a weak transverse field we obtain again a classical Ising model which includes also next-to-nearest neighborg:
\begin{align}
	\tilde{H} = K \sum_n \sigma_n^z \sigma_{n+1}^z + B \sum_n \sigma_n^z \sigma_{n+2}^z
	\label{H_tilde_NN}
\end{align}
where~\cite{ScalingHeyl}:
\begin{equation}
\begin{split}
K = \frac{1-\cos(4tJ)}{4J} h + iJt, \\
B = -i\frac{ht}{2} + i \frac{h}{8J}\sin(4tJ).
\end{split}
\label{BK}
\end{equation}

\subsection{Effective classical Ising model for both perturbations to the initial state and final Hamiltonian}

Combining this cumulant expansion with the results of the perturbed initial condition derived before we again find that the Loschmidt amplitude $\mathcal{G}(t)$ is given by $\tilde{H}$ in Eq.~(\ref{H_NNN}) to first order in the perturbation strengths $J_0/h_0$ and $h/J$.
To see this, let us write
\begin{equation}
	|\psi'(t)\rangle = e^{-iHt} |\psi'\rangle = e^{-iHt} e^{-S} |\psi\rangle \, .
\end{equation}
Commuting the SW transformation $e^{-S}$ past the time-evolution operator yields:
\begin{equation}
	|\psi'(t)\rangle = e^{-S} e^{-it e^{S} H e^{-S}} |\psi\rangle \, .
\end{equation}
To lowest order in $J_0/h_0$ and $h/J$ we get that
\begin{equation}
	 e^{S} H e^{-S} = H + \mathcal{O}\left( \frac{hJ_0}{Jh_0} \right).
\end{equation}
Using this result, we obtain that $\mathcal{G}(t) = \langle \psi'|\psi'(t)\rangle$ reduces to the same result we had in the Eq.~(\ref{H_NNN}).

\subsection{Renormalization group analysis}

In the following, we study the influence of the longer-ranged couplings using RG arguments, which are different from the ones used previously~\cite{sachdev2007,stanley1999scaling}.
Let us introduce bond spin operators~\cite{nelson1975} defined as $\tau_n = \sigma_n^z\sigma_{n+1}^z$.
Then, it is possible to recast the Hamiltonian in \eqref{H_tilde_NN} as a classical Ising Hamiltonian with only nearest neighbor interactions and a longitudinal field \cite{nelson1975} and Eq. \eqref{H_tilde_NN} becomes:
\begin{equation}
\tilde{H}_\tau = K \sum_n \tau_n^z + B \sum_n \tau_n^z \tau_{n+1}^z.
\label{tau_H}
\end{equation}
This map is exact, but special care is needed when considering the partition function, and thus the Loschmidt amplitude, since $\mathrm{Tr}_\sigma e^{\tilde H} \neq \mathrm{Tr}_\tau e^{\tilde{H}_\tau}$.
Here, $\mathrm{Tr}_\sigma$ refers to a trace over all spin configurations of the original $\sigma_n^z$ Pauli matrices ($\sigma$-spin basis) whereas $\mathrm{Tr}_\tau$ to the trace over all configurations of the bond spin operators ($\tau$-spin basis).
The reason for the difference between the traces over the different sets of configurations is the $\mathbb{Z}_2$ symmetry  in the Hamiltonian $\tilde H$ which can be cast in the form $\sigma^z_n \rightarrow -\sigma^z_n$.

Let us give a brief example to better understand the origin of the inequality of the traces: $\mathrm{Tr}_\sigma e^{\tilde H} \neq \mathrm{Tr}_\tau e^{\tilde{H}_\tau}$, although the existence of an exact map between the two Hamiltonians.
For simplicity we consider a chain with $N=4$ lattice sites described by the Hamiltonian
\begin{equation}
 H_{\sigma} = -\sum_{n=1}^4 \sigma_n^z \sigma_{n+1}^z,
 \label{H_example_1}
\end{equation}
with periodic boundary conditions. According to the exact map, the corresponding $H_{\tau}$ reads:
\begin{equation}
H_{\tau} = -\sum_{n=1}^4 \tau_n.
\label{H_example_2}
\end{equation}
When computing the partition function $Z_{\sigma}=\text{Tr}_{\sigma}\;e^{H_{\sigma}}$ using the $\sigma$-spin basis, the trace becomes a classical object equals to $\sum_s e^{\sum_n s_n s_{n+1}}$, where $s_n$ is eigenvalue of $\sigma_n^z$. One can easily realize that for each configuration $s$, the outcome of the sum $\sum_{n=1}^4 s_n s_{n+1}$ can either be $\pm 2$ or $0$.
On the other hand, using naively the map to the bond operators and considering the partition function  $Z_{\tau}=\text{Tr}_{\tau}\;e^{H_{\tau}} = \sum_{s^{\tau}} e^{\sum_n s^{\tau}_n}$, with $s^{\tau}_n$ is eigenvalue of $\tau_n$,  it comes out that the sum $\sum_{n=1}^4 s^{\tau}_n$ can assume the values $ \pm 2, \pm 1, 0$ leading to a different result for the two partition functions $Z_{\sigma}$ and $Z_{\tau}$. 
Generalizing what observe to the case with $N=2^n, \; \; n \in \mathbb{N}$ it is still possible to use the bond map to compute $Z_{\sigma}$, limiting the trace in $Z_{\tau}$ over the $\tau$-spin configurations
whose magnetization $m_{\tau}= \sum_{n} s_n^{\tau}$ is an even number. This effectively means that the number of domain walls in the spin configurations for the $\sigma$ operators is always an even number for periodic boundary conditions that we use. Let us define a projector $P_e$ onto the subspace of an even number of domain walls:
\begin{equation}
	P_e = e^{-i\pi \sum_{n}\tau_n}+1 \, ,
\end{equation}
Using $P_e$ we can find a general relation between the two partition functions:
\begin{equation}
\text{Tr}_\sigma\;e^{\tilde H} =\text{Tr}_\tau \left[ \; \left(e^{-i\pi \sum_{n}\tau_n}+1 \right)e^{\tilde{H}_\tau} \right].
\label{partition}
\end{equation}
The right-hand side of Eq. \eqref{partition} can be decomposed into the sum of two partition functions of the Hamiltonians $\tilde{H}_1$ and $\tilde{H}_2$ both given by \eqref{tau_H}, but one of the two (let's say $\tilde{H}_2$ for instance) has an additional longitudinal field equal to $-i\frac{\pi}{2}$:
\begin{equation}
\text{Tr}_\sigma\;e^{H(\sigma)} = \text{Tr}_\tau \; e^{\tilde{H}_1(\tau)} + \text{Tr}_\tau \; e^{\tilde{H}_2(\tau)},
\label{partition_2}
\end{equation}
with
\begin{equation}
	\begin{split}
	\tilde{H}_1 = K \sum_n \tau_n^z + B \sum_n \tau_n^z \tau_{n+1}^z,\\
	\tilde{H}_2 = (K-i\frac{\pi}{2}) \sum_n \tau_n^z + B \sum_n \tau_n^z \tau_{n+1}^z.
	\label{tau_H_1}
	\end{split}
\end{equation}
In the end we have now mapped the Loschmidt amplitude onto the sum of two partition functions for the bond spin operators.
These two partition functions admit an exact RG transformation which was extensively discussed in Sec.~\ref{SBP}.

Summarizing, we find that after performing the exact RG transformation the resulting Ising model for the bond spin operators incorporates only a renormalized longitudinal field with the initial spin couplings flowing to zero.
Interpreting this result in terms of the original spin degrees of freedom this fixed point Hamiltonian is equivalent to a nearest-neighbor Ising chain with renormalized couplings.
Thus, the transverse field constitutes an irrelevant perturbation in the RG sense leaving the universal features of the DQPT invariant.

\section{Magnetization fluctuations}

\begin{figure}[tb!]
	\includegraphics[width=0.95\columnwidth]{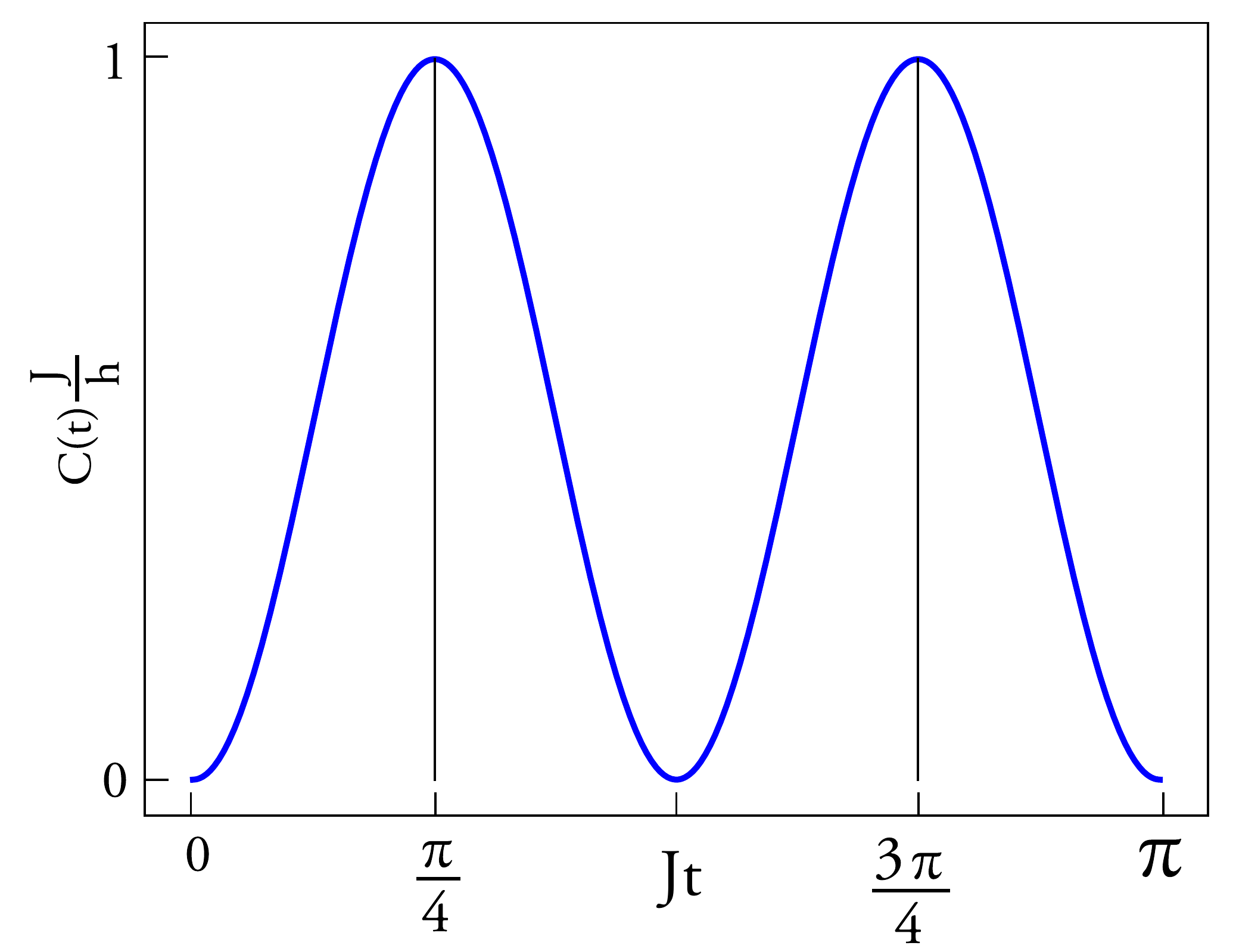}
	\caption{Fluctuations of the magnetization $C(t)$ as a function of $Jt$. The result has been normalized in such a way that the maximum value is equal to one. We observe that the curve exhibits its maximum when a DQPT occurs: at $t=\pi/4$ and $t=3\pi/4$. }
	\label{fig:fluctuation_magnetization}
\end{figure}

From the RG analysis the DQPT is associated with a divergent correlation length.
In the following we aim to show that this becomes manifest in the dynamics of the magnetization fluctuations:
\begin{equation}
C(t) = \frac{1}{N} \langle \psi(t)| \mathcal{M}^2 | \psi(t)\rangle\,, \quad \mathcal{M}=\sum_{n} \sigma_n^z,
\label{C_t}
\end{equation}
which can also be recast as:
\begin{equation}
\begin{split}
C(t) = \frac{1}{N}\sum_{r,l} C_{rl}(t), \quad C_{rl}(t) = \langle \psi(t) | \sigma^z_r \sigma^z_l |\psi(t)\rangle.
\end{split}
\label{C_sum}
\end{equation}
For the nearest-neighbor $C_{r,r+1}(t)$ such a relation has already been found in recent works~\cite{ScalingHeyl,schmitt2017quantum}. 
We compute $C(t)$ for the general scenario of nonvanishing transverse fields $0<h\ll J$ in the final Hamiltonian discussed in Sec.~\ref{SPP}, since $h=0$ represents a singular limit. 
There, $\mathcal{M}(t) =\mathcal{M}$ is a constant of motion.
Importantly, the nature of the DQPT for  $0<h\ll J$ is the same as for the $h=0$ case, as discussed in Sec.~\ref{SPP}.
As will be derived directly afterwards, to leading order in $h/J$ we find the following result:
\begin{equation}
\begin{split}
C(t)  = \frac{h}{2J}[ 1-\cos (4Jt)] \,,
\end{split}
\label{fluct_main}
\end{equation}
shown also in Fig.~\ref{fig:fluctuation_magnetization}.
For the nearest-neighbor correlations similar observations have already been made recently~\cite{ScalingHeyl,schmitt2017quantum}.
The above result can be obtained based on the same techniques as in Sec.~\ref{SPP}. In particular, moving to interaction picture we are allowed to express the time-evolved state $|\psi(t)\rangle$  in terms of a classical network of Ising spins as in Eq.~\eqref{psi_cumulant}, which leads to the following form for $C_{rl}(t)$:
\begin{equation}
C_{rl}(t) = \sum_{s,s'}e^{it(E_{s'}-E_s)} \langle \psi| W^{\dagger}(t) |s'\rangle \langle s'| \sigma^z_r \sigma^z_l | s\rangle \langle s| W(t)|\psi\rangle,
\label{C_rl}
\end{equation}
where
\begin{equation}
\begin{split}
	W(t) = \mathcal{T}\exp\left( -i \int_0^t dt' \lambda V(t')\right), \\ \quad \lambda V(t) =e^{iH_0t}Ve^{-iH_0t},\\ \quad H_0 = -J \sum_n \sigma_n^z \sigma_{n+1}^z, \quad V =-h \sum_n \sigma_n^x.
\end{split}
\end{equation}
Noting that $\langle s'| \sigma^z_r \sigma^z_l | s\rangle = s_r s_l \delta_{s,s'} $, one of the two sums in Eq.~\eqref{C_rl} vanishes.
Moreover, in the limit $h \ll J$, we approximate $ \langle s| W(t)|\psi\rangle$ appearing in Eq.~\eqref{C_rl} to first order in $h/J$ according to Eq.~\eqref{Cumulant}.
As a consequence, we obtain as a final result that $C_{rl}(t)$ assumes a form analogous of the two-point correlation function at equilibrium:
\begin{equation}
\begin{split}
C_{rl}(t) = 
\frac{\sum_s  e^{\frac{h}{2J}[1-\cos(4Jt)]\sum_n s_n s_{n+1}}  s_r s_l}{\sum_s e^{\frac{h}{2J}[1-\cos(4Jt)]\sum_n s_n s_{n+1}}}
= \frac{\text{Tr} \left[ e^\mathcal{H}   \sigma_r^z \sigma_l^z \right]}{\text{Tr} \left[ e^\mathcal{H} \right]},
\end{split}
\label{fluct_2}
\end{equation}
where $ \mathcal{H} = \frac{h}{2J}[1-\cos(4Jt)]\sum_n \sigma^z_n \sigma^z_{n+1}$ and $s_i$ is the eigenvalue of $\sigma^z_i$.
Since the above expression is a conventional equilibrium correlation function, we can adopt transfer matrix techniques~\cite{sachdev2007} to obtain as the final result for the fluctuations of the magnetization $C(t)$ in Eq.~\eqref{C_sum}:
\begin{equation}
\begin{split}
C(t) = \frac{\tanh \left( \frac{h}{2J}[ 1-\cos (4Jt)] \right) }{1- \tanh \left( \frac{h}{2J}[ 1-\cos (4Jt)] \right) }  
\simeq \frac{h}{2J}[ 1-\cos (4Jt)],
\end{split}
\label{fluct_3}
\end{equation}
where the approximation in Eq. \eqref{fluct_3} holds in the limit $h \ll J$.

As one can see from Fig.~\ref{fig:fluctuation_magnetization}, the magnetization fluctuations become maximal at those times where DQPTs occur.
This observation might already find a convincing explanation by the divergent correlation length of the underlying DQPT. 
Alternatively, one might adopt an  interpretation of DQPTs as a dynamical analog of conventional quantum phase transitions which has been outlined in Ref.~\cite{heyl2014dynamical,Heyl2017Review} and also experimentally measured~\cite{jurcevic2016}.
The Loschmidt amplitude $\mathcal{G}(t)$ is a projection of the time-evolved state back onto the ground state of the initial Hamiltonian and in this way it quantifies the dynamics in its ground state manifold, which can be thought of as a nonequilibrium equivalent to zero temperature.
From this perspective, we observe from the theory of DQPTs that the magnetization fluctuations diverge at the critical times in the ground state manifold.
Observables or correlation functions, however, acquire their dominant contribution from elevated energy densities beyond the ground state due to the energy pumped into the system by the quantum quench.
Interpreting these excited energy densities as the nonequilibrium counterpart to temperature, it is natural to expect that the divergent magnetization correlations are then cut off at elevated energy densities making the magnetization fluctuations finite.
However, one can still observe the influence of the underlying DQPT through the maxima of $C(t)$ at the critical times.

\section{Conclusions} \label{Conclusions}
To further investigate the analogies between DQPTs and conventional phase transitions, we introduce in this work a dynamical analogue to a free energy density by using the formal equivalence between Loschmidt amplitudes and classical partition functions for the 1D transverse field Ising model.
While in equilibrium systems the free energy is a real-valued function, the corresponding quantity in the nonequilibrium regime becomes complex. As a consequence we observe that the conventional organization principle of free energy minimization is transformed into the more general form of a saddle-point principle. 
Moreover, we find that the effective free energy admits an expansion as a function of the complex-valued magnetization, which plays the role of the order parameter, in the spirit of the conventional Landau theory.
Quartic terms always appear to be subleading which we have traced back to the property of the one-dimensional classical Ising chain, in that it exhibits symmetry-breaking only at zero temperature and therefore at a singular point.
This makes a study in higher dimensions, e.g. for the two-dimensional Ising model, where the system exhibits an extended symmetry-broken phase, particularly promising.
Importantly, the presented formalism of the construction of the effective free energy is independent of dimensionality and can therefore be directly applied also, for example, to the two-dimensional case.
A further interesting aspect in the future is to which extent analogous effective free energies can be formulated for other systems beyond Ising models.
The formalism presented in this manuscript requires a classical limit of the Loschmidt amplitude, meaning that it can be expressed as a complex partition function of an effective classical Hamiltonian. To which extend such a mapping is possible on general grounds for other models than the Ising systems studied here, remains to be addressed in the future.

\begin{acknowledgements}
	This work was supported by the  Deutsche  Forschungsgemeinschaft  via  the  Gottfried
	Wilhelm Leibniz Prize program.
\end{acknowledgements}

\appendix
\section{magnetization density} \label{Appendix}
The equation of state~\eqref {SPA_T} can be computed exactly in the thermodynamic limit $N \rightarrow \infty$. Let us recall the EoS which relates the magnetization density $m$ and the longitudinal field $\mu$:
\begin{equation}
    mN = \frac{1}{G(\mu,t)} Tr \sum_n[ s^z_n  e^{\mathcal{H}(s,t) - \mu \sum_l s_l}] \, .
    \label{SPA}
\end{equation}
As expected, Eq.~\eqref{SPA} says that the magnetization density is given by the mean value of spins $s_z$.
Using transfer matrix techniques, it is possible to solve exactly both the numerator and the denominator of Eq. \eqref {SPA}. The latter was already obtained in Eq.~\eqref{G(mu)}, while we now aim to compute the former:
\begin{equation}
 \begin{split}
Tr \sum_n[ s^z_n  e^{\mathcal{H} - \mu \sum_l s_l}] = Tr[\sigma_zD^N] 
  = N [\Gamma_1 \varepsilon_1^N + \Gamma_2\varepsilon_2^N] \, ,
 \end{split}
 \label{EoS_app}
\end{equation}
where
\begin{equation}
\begin{aligned}
 \Gamma_1 = \frac{(1-|\alpha_1|^2)(|a|^2+|c|^2)}{1+|\alpha_1|^2}  + \frac{(1-\alpha_1^*\alpha_2)(b^*a+d^*c)}{\sqrt{1+|\alpha_1|^2}\sqrt{1+|\alpha_2|^2}}, \\
 \Gamma_2 = \frac{(1-|\alpha_2|^2)(|b|^2+|d|^2)}{1+|\alpha_2|^2}  + \frac{(1-\alpha_2^*\alpha_1)(a^*b+c^*d)}{\sqrt{1+|\alpha_1|^2}\sqrt{1+|\alpha_2|^2}}, \\
 \end{aligned}
\end{equation}
 \begin{equation}
\begin{aligned}
 \alpha_{1(2)} = -x^{-1} (x^{-1}y^{-1}-\varepsilon_{1(2)}) \, , \quad x = e^{K},\quad  y = e^{B}.
 \end{aligned}
 \label{alpha}
\end{equation}
The parameters $a,b,c,d$ given by:
 \begin{equation}
 \begin{aligned}
 a =\frac{v_2(2)}{\text{det}(v)}\, , \quad
 b =\frac{-v_1(2)}{\text{det}(v)}\, , \quad
 c =\frac{-v_2(1)}{\text{det}(v)}\, , \quad
 d =\frac{v_1(1)}{\text{det}(v)} \, ,
 \end{aligned}
 \label{abcd}
 \end{equation}
are well defined as long as $\text{det}(v) \neq 0$ where $\text{det}(v)$ stands for the determinant.
Since $v$ is the matrix containing the right eigenvectors $v_1, \;v_2$ of the matrix $T$ introduced in Eq. \eqref{D}, the requirement $\text{det}(v) \neq 0$ is equivalent to ask that $v_1, \;v_2$ are not parallel. These eigenvectors are given by:

 \begin{equation}
\begin{aligned}
   v_1 = \frac{1}{\sqrt{1+|\alpha_1|^2}} \left( \begin{array}{c}1  \\
\alpha_1  \end{array} \right) \, , \quad
  v_2 = \frac{1}{\sqrt{1+|\alpha_2|^2}} \left( \begin{array}{c}1  \\
\alpha_2  \end{array} \right). \\
\end{aligned}
\end{equation}
From Eqs.~\eqref{EoS_app},~\eqref{G(mu)}, we can extract the behavior of the magnetization density $m$ in the limit $N \rightarrow \infty$ where assumes the following form:
\begin{equation}
    m = \Gamma,
    \label{SPA_density}
\end{equation}
where $\Gamma = \Gamma_{1/2}$ according to $\varepsilon = \varepsilon_{1/2}$.

\section{range of validity of Landau theory}
\begin{figure}[tb]
	\includegraphics[width=1\columnwidth]{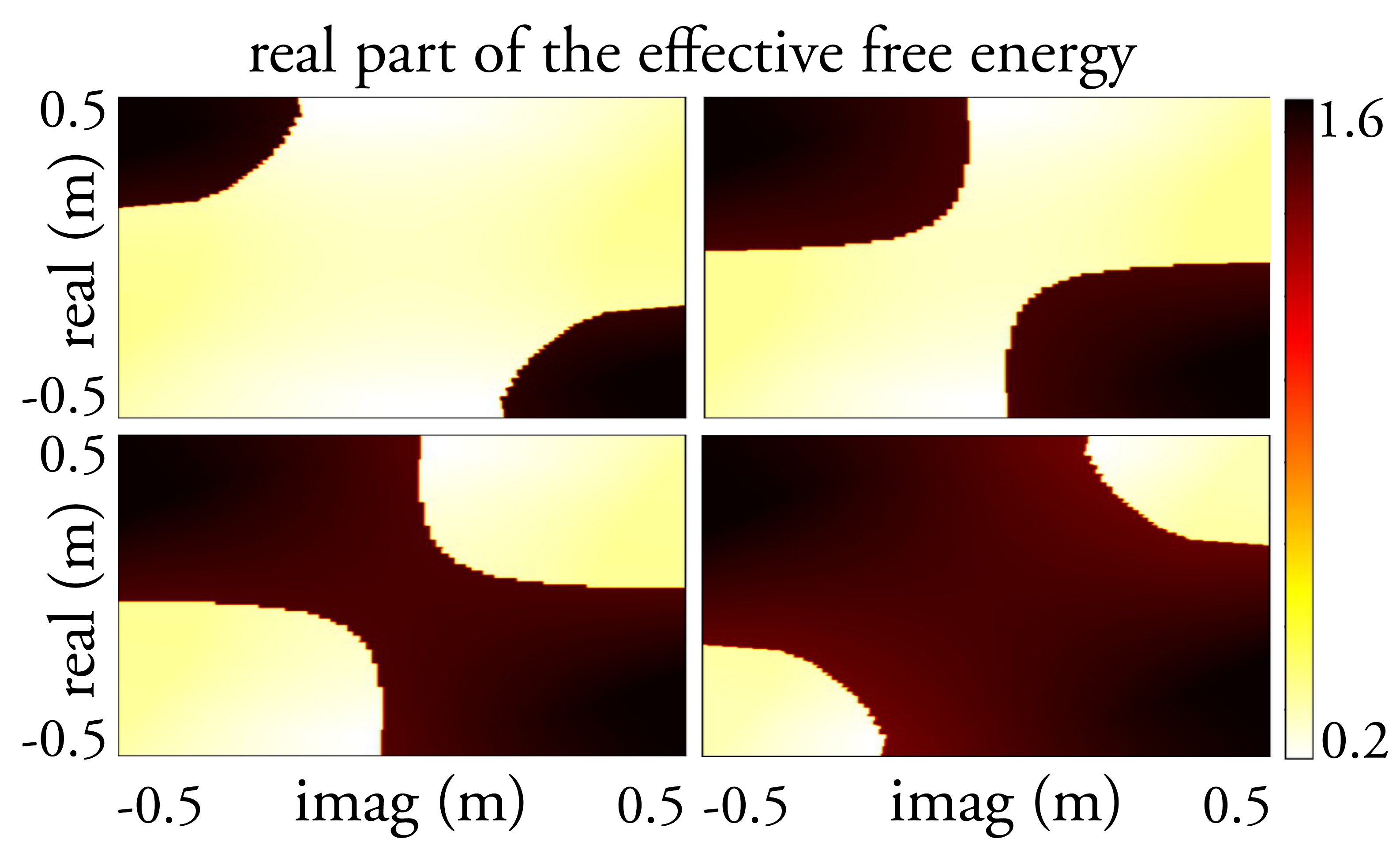}
	\caption{Absolute value of the effective free energy $f(m,t)$ for the following values of $Jt$: first row $Jt=\pi\frac{73}{320}$ and $Jt=\pi\frac{79}{320}$. Second row: $Jt=\pi\frac{81}{320}$ and $Jt=\pi\frac{87}{320}$ . The jumps of the effective free energy $f(m,t)$ are manifested in the plot as sudden changes of the colors dark red and yellow.}
	\label{fig:jumps_free_en}
\end{figure}

The polynomial expansion of the effective free energy in terms of the complex magnetization $m$ shown in Eq. \eqref{fit_f_m} is valid as long as the values of $m$
are small enough to guarantee that the effective free energy does not show any jumps. The domain of validity depends on time, in particular it becomes
smaller and smaller as $t$ approaches $t_c$. This is shown in Fig. \ref{fig:jumps_free_en}, where the absolute value of the effective free energy has some jumps and their location is more and more close to the origin as the time flows toward the critical time. Exactly at that time, the domain of validity has zero measure.
The origin of these jumps is due to the change of the dominant eigenvalue $\varepsilon_{1/2}$ appearing in Eq. \eqref{G(mu)}. 
 Expanding the effective free energy in terms of $m$ in the vicinity of the origin, the magnetization density can be neglected compare to $g(h)$ and according to Eq. \eqref{LT} the effective free energy can be approximated as $ f(m) \sim g(h) = -\text{log}(\varepsilon)$. 
From this formula we can directly observe that the nonanalytic behavior of $\varepsilon$ is then reflected onto the effective free energy.

\section{Scaling of the effective free energy close to the critical time}
\begin{figure}[tb!]
	\includegraphics[width=1.\columnwidth]{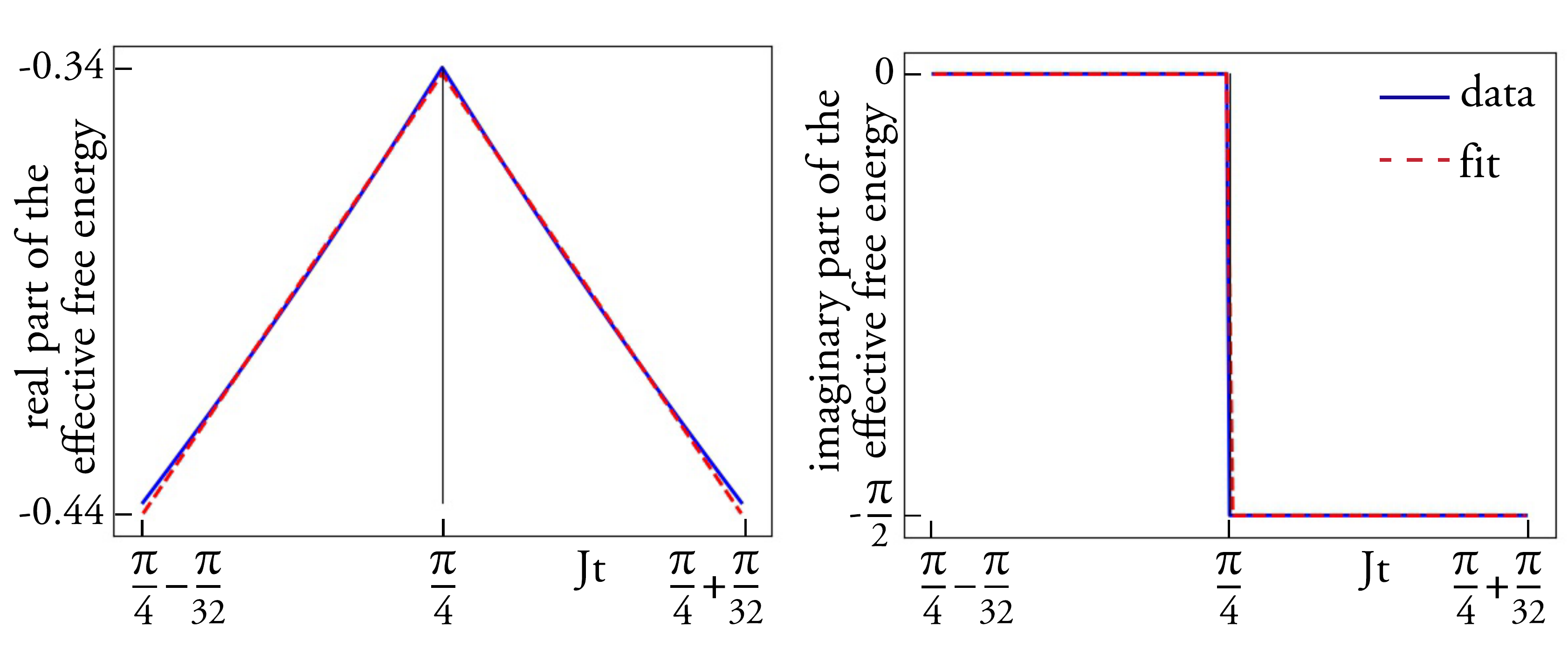}\hfil
	\caption{Left: comparison between the exact result (blue continuous line) and the one obtained through the expansion~\eqref{fit_f_m} (red dotted line) of the  real part of the dominant effective free energy~\eqref{f_saddle_point}.
 Right: same analysis but of the imaginary part.}
	\label{fig:fit_loschmidt}
\end{figure}

To test the validity of our result, we compare the dominant effective free energy $f(t)$ in Eq.~\eqref{f_saddle_point} computed through the exact result and with the polynomial expression in Eq.~\eqref{fit_f_m}. 
Fig.~\ref{fig:fit_loschmidt} shows in blue the exact result corresponding to Eq~\eqref{f_saddle_point} where $f(m,t)$ is given by Eq.~\eqref{eq_fm_sp}, while the curve coming from the expansion \eqref{fit_f_m},  is shown in red to leading term in N.
In the limit $N\rightarrow \infty$, according to Eq.~\eqref{f_saddle_point} the dominant contribution of $f(m,t)$ comes from $f_0$ since the saddle point condition is fulfill at vanishing magnetization density $\tilde{m}=0$, therefore the constant term $f_0(t)$ in the expansion \eqref{fit_f_m} is given by
\begin{equation}
f_0(t) = f(m=0,t).
\end{equation}
Moreover, Fig.~\ref{fig:fit_loschmidt} suggests the possibility to approximate $\mathrm{Re} [ f_0 ]$ with a linear expression and $\mathrm{Im} [ f _0]$ with a Heaviside theta function.
In particular,
$ \mathrm{Re} [ f_0 ]= \theta \delta+\tau$ where $\theta$ and $\delta$ take care respectively the change
of sign and the value of the slope, while $\tau$ provides a vertical shift of the curve. 
From the fitting we find that $\delta \simeq 0.92$, $\tau \simeq -1$ and $\theta = \vert \frac{Jt_c - Jt}{Jt_c} \vert$ as announced in Eq.~\eqref{theta} . We can conclude that, close to the critical time,  the real part of the free energy scales linearly like: $ \mathrm{Re} [ f ] \sim - \vert \frac{Jt_c - Jt}{Jt_c} \vert$.
As concern the imaginary part, 
we have that
$\mathrm{Im} [ f_0 ]= -\frac{\pi}{4} -\text{sign}(\theta)\frac{\pi}{4}$,
describing the sudden jump at the critical time.

\begin{figure}[tb!]
	\includegraphics[width=1.02\columnwidth]{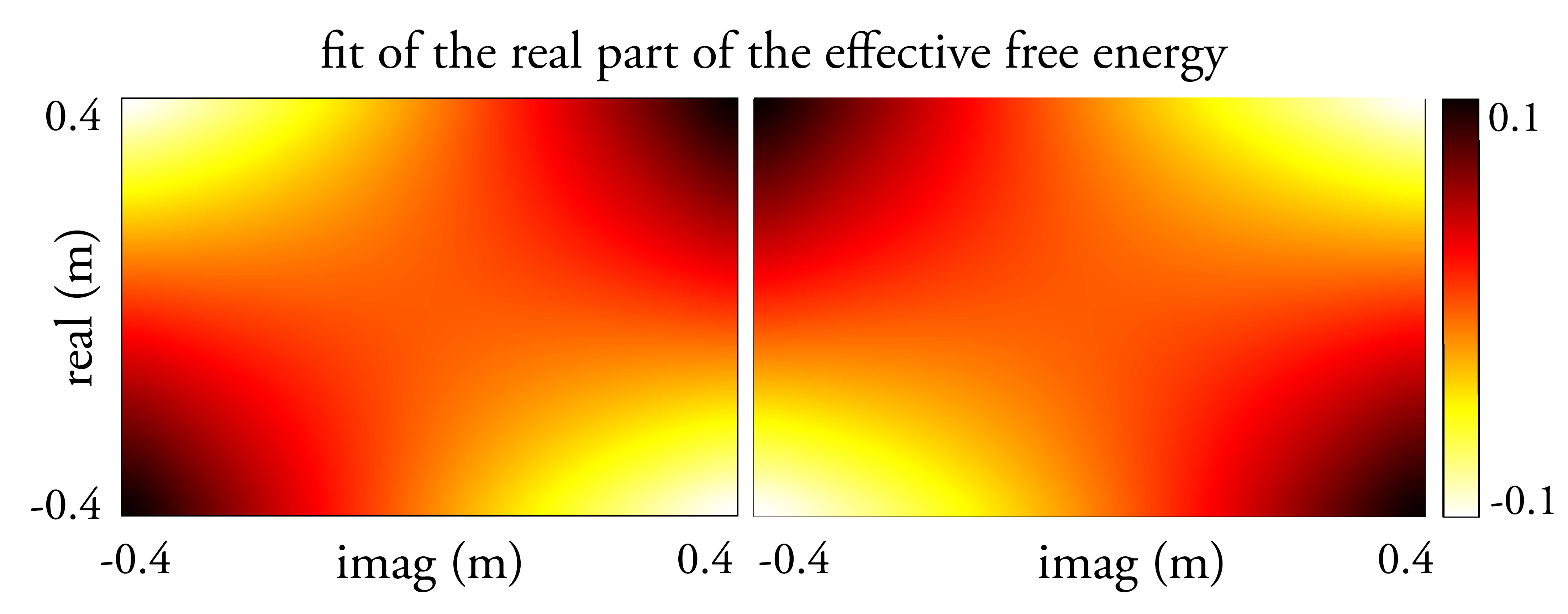} \hfil
	\caption{Real part of the expansion of effective free energy density given by Eq~\eqref{fit_f_m} at different times in the complex magnetization density plane.
		The times chosen and the interval of the complex magnetization density $m$ are the same as the ones of Fig.~\ref{fig:NO_jumps_real_free} to allow a direct comparison with that figure. The value of $\alpha$ coming from the best fitting procedure is: $\alpha = 0.98+0.5i$. }
	\label{fig:NO_jumps_real_free_fit}
\end{figure}

Another evidence supporting the validity of the polynomial expansion of the effective free energy in Eq.~\eqref{fit_f_m}, is the similiarity between
Fig.~\ref{fig:NO_jumps_real_free_fit}, which shows the effective free energy computed according to the Eq.~\eqref{fit_f_m} in the complex magnetization plane 
and Fig. \ref{fig:NO_jumps_real_free}, where the exact effective free energy given by Eq.~\eqref{eq_fm_sp} is shown.
The parameter $\alpha$ appearing in Eq.~\eqref{fit_f_m} is obtained from fitting procedure, i.e. we minimize the difference between the exact formula~\eqref{eq_fm_sp} and the expansion~\eqref{fit_f_m}. In the end we get $\alpha \simeq 0.98+0.5i $.

\bibliography{BIB_1}

\begin{thebibliography}{45}%
\makeatletter
\providecommand \@ifxundefined [1]{%
 \@ifx{#1\undefined}
}%
\providecommand \@ifnum [1]{%
 \ifnum #1\expandafter \@firstoftwo
 \else \expandafter \@secondoftwo
 \fi
}%
\providecommand \@ifx [1]{%
 \ifx #1\expandafter \@firstoftwo
 \else \expandafter \@secondoftwo
 \fi
}%
\providecommand \natexlab [1]{#1}%
\providecommand \enquote  [1]{``#1''}%
\providecommand \bibnamefont  [1]{#1}%
\providecommand \bibfnamefont [1]{#1}%
\providecommand \citenamefont [1]{#1}%
\providecommand \href@noop [0]{\@secondoftwo}%
\providecommand \href [0]{\begingroup \@sanitize@url \@href}%
\providecommand \@href[1]{\@@startlink{#1}\@@href}%
\providecommand \@@href[1]{\endgroup#1\@@endlink}%
\providecommand \@sanitize@url [0]{\catcode `\\12\catcode `\$12\catcode
  `\&12\catcode `\#12\catcode `\^12\catcode `\_12\catcode `\%12\relax}%
\providecommand \@@startlink[1]{}%
\providecommand \@@endlink[0]{}%
\providecommand \url  [0]{\begingroup\@sanitize@url \@url }%
\providecommand \@url [1]{\endgroup\@href {#1}{\urlprefix }}%
\providecommand \urlprefix  [0]{URL }%
\providecommand \Eprint [0]{\href }%
\providecommand \doibase [0]{http://dx.doi.org/}%
\providecommand \selectlanguage [0]{\@gobble}%
\providecommand \bibinfo  [0]{\@secondoftwo}%
\providecommand \bibfield  [0]{\@secondoftwo}%
\providecommand \translation [1]{[#1]}%
\providecommand \BibitemOpen [0]{}%
\providecommand \bibitemStop [0]{}%
\providecommand \bibitemNoStop [0]{.\EOS\space}%
\providecommand \EOS [0]{\spacefactor3000\relax}%
\providecommand \BibitemShut  [1]{\csname bibitem#1\endcsname}%
\let\auto@bib@innerbib\@empty
\bibitem [{\citenamefont {Sachdev}(2007)}]{sachdev2007}%
  \BibitemOpen
  \bibfield  {author} {\bibinfo {author} {\bibfnamefont {S.}~\bibnamefont
  {Sachdev}},\ }\href@noop {} {\emph {\bibinfo {title} {{Quantum phase
  transitions}}}}\ (\bibinfo  {publisher} {Wiley Online Library},\ \bibinfo
  {year} {2007})\BibitemShut {NoStop}%
\bibitem [{\citenamefont {Huang}(2009)}]{huang2009}%
  \BibitemOpen
  \bibfield  {author} {\bibinfo {author} {\bibfnamefont {K.}~\bibnamefont
  {Huang}},\ }\href@noop {} {\emph {\bibinfo {title} {{Introduction to
  statistical physics}}}}\ (\bibinfo  {publisher} {CRC Press},\ \bibinfo {year}
  {2009})\BibitemShut {NoStop}%
\bibitem [{\citenamefont {Heyl}\ \emph {et~al.}(2013)\citenamefont {Heyl},
  \citenamefont {Polkovnikov},\ and\ \citenamefont {Kehrein}}]{heyl2013}%
  \BibitemOpen
  \bibfield  {author} {\bibinfo {author} {\bibfnamefont {M.}~\bibnamefont
  {Heyl}}, \bibinfo {author} {\bibfnamefont {A.}~\bibnamefont {Polkovnikov}}, \
  and\ \bibinfo {author} {\bibfnamefont {S.}~\bibnamefont {Kehrein}},\ }\href
  {\doibase 10.1103/PhysRevLett.110.135704} {\bibfield  {journal} {\bibinfo
  {journal} {Phys. Rev. Lett.}\ }\textbf {\bibinfo {volume} {110}},\ \bibinfo
  {pages} {135704} (\bibinfo {year} {2013})}\BibitemShut {NoStop}%
\bibitem [{\citenamefont {Zvyagin}(2016)}]{Zvyagin2017Review}%
  \BibitemOpen
  \bibfield  {author} {\bibinfo {author} {\bibfnamefont {A.~A.}\ \bibnamefont
  {Zvyagin}},\ }\href {\doibase 10.1063/1.4969869} {\bibfield  {journal}
  {\bibinfo  {journal} {Low Temperature Physics}\ }\textbf {\bibinfo {volume}
  {42}},\ \bibinfo {pages} {971} (\bibinfo {year} {2016})}\BibitemShut
  {NoStop}%
\bibitem [{\citenamefont {Heyl}(2018)}]{Heyl2017Review}%
  \BibitemOpen
  \bibfield  {author} {\bibinfo {author} {\bibfnamefont {M.}~\bibnamefont
  {Heyl}},\ }\href {http://stacks.iop.org/0034-4885/81/i=5/a=054001} {\bibfield
   {journal} {\bibinfo  {journal} {Reports on Progress in Physics}\ }\textbf
  {\bibinfo {volume} {81}},\ \bibinfo {pages} {054001} (\bibinfo {year}
  {2018})}\BibitemShut {NoStop}%
\bibitem [{\citenamefont {{Fl{\"a}schner N.}}\ \emph
  {et~al.}(2017)\citenamefont {{Fl{\"a}schner N.}}, \citenamefont {{Vogel D.}},
  \citenamefont {{Tarnowski M.}}, \citenamefont {{Rem B. S.}}, \citenamefont
  {{L{\"u}hmann D.-S.}}, \citenamefont {{Heyl M.}}, \citenamefont {{Budich J.
  C.}}, \citenamefont {{Mathey L.}}, \citenamefont {{Sengstock K.}},\ and\
  \citenamefont {{Weitenberg C.}}}]{Flaeschner2017}%
  \BibitemOpen
  \bibfield  {author} {\bibinfo {author} {\bibnamefont {{Fl{\"a}schner N.}}},
  \bibinfo {author} {\bibnamefont {{Vogel D.}}}, \bibinfo {author}
  {\bibnamefont {{Tarnowski M.}}}, \bibinfo {author} {\bibnamefont {{Rem B.
  S.}}}, \bibinfo {author} {\bibnamefont {{L{\"u}hmann D.-S.}}}, \bibinfo
  {author} {\bibnamefont {{Heyl M.}}}, \bibinfo {author} {\bibnamefont {{Budich
  J. C.}}}, \bibinfo {author} {\bibnamefont {{Mathey L.}}}, \bibinfo {author}
  {\bibnamefont {{Sengstock K.}}}, \ and\ \bibinfo {author} {\bibnamefont
  {{Weitenberg C.}}},\ }\href {\doibase 10.1038/s41567-017-0013-8} {\bibfield
  {journal} {\bibinfo  {journal} {Nature Phys.}\ } (\bibinfo {year} {2017}),\
  10.1038/s41567-017-0013-8}\BibitemShut {NoStop}%
\bibitem [{\citenamefont {Jurcevic}\ \emph {et~al.}(2017)\citenamefont
  {Jurcevic}, \citenamefont {Shen}, \citenamefont {Hauke}, \citenamefont
  {Maier}, \citenamefont {Brydges}, \citenamefont {Hempel}, \citenamefont
  {Lanyon}, \citenamefont {Heyl}, \citenamefont {Blatt},\ and\ \citenamefont
  {Roos}}]{jurcevic2016}%
  \BibitemOpen
  \bibfield  {author} {\bibinfo {author} {\bibfnamefont {P.}~\bibnamefont
  {Jurcevic}}, \bibinfo {author} {\bibfnamefont {H.}~\bibnamefont {Shen}},
  \bibinfo {author} {\bibfnamefont {P.}~\bibnamefont {Hauke}}, \bibinfo
  {author} {\bibfnamefont {C.}~\bibnamefont {Maier}}, \bibinfo {author}
  {\bibfnamefont {T.}~\bibnamefont {Brydges}}, \bibinfo {author} {\bibfnamefont
  {C.}~\bibnamefont {Hempel}}, \bibinfo {author} {\bibfnamefont {B.~P.}\
  \bibnamefont {Lanyon}}, \bibinfo {author} {\bibfnamefont {M.}~\bibnamefont
  {Heyl}}, \bibinfo {author} {\bibfnamefont {R.}~\bibnamefont {Blatt}}, \ and\
  \bibinfo {author} {\bibfnamefont {C.~F.}\ \bibnamefont {Roos}},\ }\href
  {\doibase 10.1103/PhysRevLett.119.080501} {\bibfield  {journal} {\bibinfo
  {journal} {Phys. Rev. Lett.}\ }\textbf {\bibinfo {volume} {119}},\ \bibinfo
  {pages} {080501} (\bibinfo {year} {2017})}\BibitemShut {NoStop}%
\bibitem [{\citenamefont {Heyl}(2015)}]{ScalingHeyl}%
  \BibitemOpen
  \bibfield  {author} {\bibinfo {author} {\bibfnamefont {M.}~\bibnamefont
  {Heyl}},\ }\href {\doibase 10.1103/PhysRevLett.115.140602} {\bibfield
  {journal} {\bibinfo  {journal} {Phys. Rev. Lett.}\ }\textbf {\bibinfo
  {volume} {115}},\ \bibinfo {pages} {140602} (\bibinfo {year}
  {2015})}\BibitemShut {NoStop}%
\bibitem [{\citenamefont {Sharma}\ \emph {et~al.}(2015)\citenamefont {Sharma},
  \citenamefont {Suzuki},\ and\ \citenamefont {Dutta}}]{sharma2015quenches}%
  \BibitemOpen
  \bibfield  {author} {\bibinfo {author} {\bibfnamefont {S.}~\bibnamefont
  {Sharma}}, \bibinfo {author} {\bibfnamefont {S.}~\bibnamefont {Suzuki}}, \
  and\ \bibinfo {author} {\bibfnamefont {A.}~\bibnamefont {Dutta}},\ }\href
  {\doibase 10.1103/PhysRevB.92.104306} {\bibfield  {journal} {\bibinfo
  {journal} {Phys. Rev. B}\ }\textbf {\bibinfo {volume} {92}},\ \bibinfo
  {pages} {104306} (\bibinfo {year} {2015})}\BibitemShut {NoStop}%
\bibitem [{\citenamefont {Kriel}\ \emph {et~al.}(2014)\citenamefont {Kriel},
  \citenamefont {Karrasch},\ and\ \citenamefont
  {Kehrein}}]{kriel2014dynamical}%
  \BibitemOpen
  \bibfield  {author} {\bibinfo {author} {\bibfnamefont {J.~N.}\ \bibnamefont
  {Kriel}}, \bibinfo {author} {\bibfnamefont {C.}~\bibnamefont {Karrasch}}, \
  and\ \bibinfo {author} {\bibfnamefont {S.}~\bibnamefont {Kehrein}},\ }\href
  {\doibase 10.1103/PhysRevB.90.125106} {\bibfield  {journal} {\bibinfo
  {journal} {Phys. Rev. B}\ }\textbf {\bibinfo {volume} {90}},\ \bibinfo
  {pages} {125106} (\bibinfo {year} {2014})}\BibitemShut {NoStop}%
\bibitem [{\citenamefont {Karrasch}\ and\ \citenamefont
  {Schuricht}(2013)}]{karrasch2013dynamical}%
  \BibitemOpen
  \bibfield  {author} {\bibinfo {author} {\bibfnamefont {C.}~\bibnamefont
  {Karrasch}}\ and\ \bibinfo {author} {\bibfnamefont {D.}~\bibnamefont
  {Schuricht}},\ }\href {\doibase 10.1103/PhysRevB.87.195104} {\bibfield
  {journal} {\bibinfo  {journal} {Phys. Rev. B}\ }\textbf {\bibinfo {volume}
  {87}},\ \bibinfo {pages} {195104} (\bibinfo {year} {2013})}\BibitemShut
  {NoStop}%
\bibitem [{\citenamefont {Budich}\ and\ \citenamefont
  {Heyl}(2016)}]{budich2016dynamical}%
  \BibitemOpen
  \bibfield  {author} {\bibinfo {author} {\bibfnamefont {J.~C.}\ \bibnamefont
  {Budich}}\ and\ \bibinfo {author} {\bibfnamefont {M.}~\bibnamefont {Heyl}},\
  }\href {\doibase 10.1103/PhysRevB.93.085416} {\bibfield  {journal} {\bibinfo
  {journal} {Phys. Rev. B}\ }\textbf {\bibinfo {volume} {93}},\ \bibinfo
  {pages} {085416} (\bibinfo {year} {2016})}\BibitemShut {NoStop}%
\bibitem [{\citenamefont {Bhattacharya}\ \emph {et~al.}(2017)\citenamefont
  {Bhattacharya}, \citenamefont {Bandopadhyay},\ and\ \citenamefont
  {Dutta}}]{bhattacharya2017mixed}%
  \BibitemOpen
  \bibfield  {author} {\bibinfo {author} {\bibfnamefont {U.}~\bibnamefont
  {Bhattacharya}}, \bibinfo {author} {\bibfnamefont {S.}~\bibnamefont
  {Bandopadhyay}}, \ and\ \bibinfo {author} {\bibfnamefont {A.}~\bibnamefont
  {Dutta}},\ }\href {\doibase 10.1103/PhysRevB.96.180303} {\bibfield  {journal}
  {\bibinfo  {journal} {Phys. Rev. B}\ }\textbf {\bibinfo {volume} {96}},\
  \bibinfo {pages} {180303} (\bibinfo {year} {2017})}\BibitemShut {NoStop}%
\bibitem [{\citenamefont {Heyl}\ and\ \citenamefont
  {Budich}(2017)}]{HeylBudich2017}%
  \BibitemOpen
  \bibfield  {author} {\bibinfo {author} {\bibfnamefont {M.}~\bibnamefont
  {Heyl}}\ and\ \bibinfo {author} {\bibfnamefont {J.~C.}\ \bibnamefont
  {Budich}},\ }\href {\doibase 10.1103/PhysRevB.96.180304} {\bibfield
  {journal} {\bibinfo  {journal} {Phys. Rev. B}\ }\textbf {\bibinfo {volume}
  {96}},\ \bibinfo {pages} {180304} (\bibinfo {year} {2017})}\BibitemShut
  {NoStop}%
\bibitem [{\citenamefont {Karrasch}\ and\ \citenamefont
  {Schuricht}(2017)}]{karrasch2017}%
  \BibitemOpen
  \bibfield  {author} {\bibinfo {author} {\bibfnamefont {C.}~\bibnamefont
  {Karrasch}}\ and\ \bibinfo {author} {\bibfnamefont {D.}~\bibnamefont
  {Schuricht}},\ }\href {\doibase 10.1103/PhysRevB.95.075143} {\bibfield
  {journal} {\bibinfo  {journal} {Phys. Rev. B}\ }\textbf {\bibinfo {volume}
  {95}},\ \bibinfo {pages} {075143} (\bibinfo {year} {2017})}\BibitemShut
  {NoStop}%
\bibitem [{\citenamefont {Schmitt}\ and\ \citenamefont
  {Heyl}(2018)}]{schmitt2017quantum}%
  \BibitemOpen
  \bibfield  {author} {\bibinfo {author} {\bibfnamefont {M.}~\bibnamefont
  {Schmitt}}\ and\ \bibinfo {author} {\bibfnamefont {M.}~\bibnamefont {Heyl}},\
  }\href {\doibase 10.21468/SciPostPhys.4.2.013} {\bibfield  {journal}
  {\bibinfo  {journal} {SciPost Phys.}\ }\textbf {\bibinfo {volume} {4}},\
  \bibinfo {pages} {013} (\bibinfo {year} {2018})}\BibitemShut {NoStop}%
\bibitem [{\citenamefont {Polkovnikov}\ \emph {et~al.}(2011)\citenamefont
  {Polkovnikov}, \citenamefont {Sengupta}, \citenamefont {Silva},\ and\
  \citenamefont {Vengalattore}}]{RevModPhys.83}%
  \BibitemOpen
  \bibfield  {author} {\bibinfo {author} {\bibfnamefont {A.}~\bibnamefont
  {Polkovnikov}}, \bibinfo {author} {\bibfnamefont {K.}~\bibnamefont
  {Sengupta}}, \bibinfo {author} {\bibfnamefont {A.}~\bibnamefont {Silva}}, \
  and\ \bibinfo {author} {\bibfnamefont {M.}~\bibnamefont {Vengalattore}},\
  }\href {\doibase 10.1103/RevModPhys.83.863} {\bibfield  {journal} {\bibinfo
  {journal} {Rev. Mod. Phys.}\ }\textbf {\bibinfo {volume} {83}},\ \bibinfo
  {pages} {863} (\bibinfo {year} {2011})}\BibitemShut {NoStop}%
\bibitem [{\citenamefont {LeClair}\ \emph {et~al.}(1995)\citenamefont
  {LeClair}, \citenamefont {Mussardo}, \citenamefont {Saleur},\ and\
  \citenamefont {Skorik}}]{leclair1995boundary}%
  \BibitemOpen
  \bibfield  {author} {\bibinfo {author} {\bibfnamefont {A.}~\bibnamefont
  {LeClair}}, \bibinfo {author} {\bibfnamefont {G.}~\bibnamefont {Mussardo}},
  \bibinfo {author} {\bibfnamefont {H.}~\bibnamefont {Saleur}}, \ and\ \bibinfo
  {author} {\bibfnamefont {S.}~\bibnamefont {Skorik}},\ }\href {\doibase
  https://doi.org/10.1016/0550-3213(95)00435-U} {\bibfield  {journal} {\bibinfo
   {journal} {Nuclear Physics B}\ }\textbf {\bibinfo {volume} {453}},\ \bibinfo
  {pages} {581 } (\bibinfo {year} {1995})}\BibitemShut {NoStop}%
\bibitem [{\citenamefont {Sharma}\ \emph {et~al.}(2016)\citenamefont {Sharma},
  \citenamefont {Divakaran}, \citenamefont {Polkovnikov},\ and\ \citenamefont
  {Dutta}}]{sharma2016slow}%
  \BibitemOpen
  \bibfield  {author} {\bibinfo {author} {\bibfnamefont {S.}~\bibnamefont
  {Sharma}}, \bibinfo {author} {\bibfnamefont {U.}~\bibnamefont {Divakaran}},
  \bibinfo {author} {\bibfnamefont {A.}~\bibnamefont {Polkovnikov}}, \ and\
  \bibinfo {author} {\bibfnamefont {A.}~\bibnamefont {Dutta}},\ }\href
  {\doibase 10.1103/PhysRevB.93.144306} {\bibfield  {journal} {\bibinfo
  {journal} {Phys. Rev. B}\ }\textbf {\bibinfo {volume} {93}},\ \bibinfo
  {pages} {144306} (\bibinfo {year} {2016})}\BibitemShut {NoStop}%
\bibitem [{\citenamefont {Bhattacharya}\ and\ \citenamefont
  {Dutta}(2017{\natexlab{a}})}]{bhattacharua2017a}%
  \BibitemOpen
  \bibfield  {author} {\bibinfo {author} {\bibfnamefont {U.}~\bibnamefont
  {Bhattacharya}}\ and\ \bibinfo {author} {\bibfnamefont {A.}~\bibnamefont
  {Dutta}},\ }\href {\doibase 10.1103/PhysRevB.96.014302} {\bibfield  {journal}
  {\bibinfo  {journal} {Phys. Rev. B}\ }\textbf {\bibinfo {volume} {96}},\
  \bibinfo {pages} {014302} (\bibinfo {year} {2017}{\natexlab{a}})}\BibitemShut
  {NoStop}%
\bibitem [{\citenamefont {Vajna}\ and\ \citenamefont
  {D\'ora}(2015)}]{vajna2015topological}%
  \BibitemOpen
  \bibfield  {author} {\bibinfo {author} {\bibfnamefont {S.}~\bibnamefont
  {Vajna}}\ and\ \bibinfo {author} {\bibfnamefont {B.}~\bibnamefont {D\'ora}},\
  }\href {\doibase 10.1103/PhysRevB.91.155127} {\bibfield  {journal} {\bibinfo
  {journal} {Phys. Rev. B}\ }\textbf {\bibinfo {volume} {91}},\ \bibinfo
  {pages} {155127} (\bibinfo {year} {2015})}\BibitemShut {NoStop}%
\bibitem [{\citenamefont {Schmitt}\ and\ \citenamefont
  {Kehrein}(2015)}]{schmitt2015dynamical}%
  \BibitemOpen
  \bibfield  {author} {\bibinfo {author} {\bibfnamefont {M.}~\bibnamefont
  {Schmitt}}\ and\ \bibinfo {author} {\bibfnamefont {S.}~\bibnamefont
  {Kehrein}},\ }\href {\doibase 10.1103/PhysRevB.92.075114} {\bibfield
  {journal} {\bibinfo  {journal} {Phys. Rev. B}\ }\textbf {\bibinfo {volume}
  {92}},\ \bibinfo {pages} {075114} (\bibinfo {year} {2015})}\BibitemShut
  {NoStop}%
\bibitem [{\citenamefont {Huang}\ and\ \citenamefont
  {Balatsky}(2016)}]{huang2016}%
  \BibitemOpen
  \bibfield  {author} {\bibinfo {author} {\bibfnamefont {Z.}~\bibnamefont
  {Huang}}\ and\ \bibinfo {author} {\bibfnamefont {A.~V.}\ \bibnamefont
  {Balatsky}},\ }\href {\doibase 10.1103/PhysRevLett.117.086802} {\bibfield
  {journal} {\bibinfo  {journal} {Phys. Rev. Lett.}\ }\textbf {\bibinfo
  {volume} {117}},\ \bibinfo {pages} {086802} (\bibinfo {year}
  {2016})}\BibitemShut {NoStop}%
\bibitem [{\citenamefont {Bhattacharya}\ and\ \citenamefont
  {Dutta}(2017{\natexlab{b}})}]{Bhattacharya2017i}%
  \BibitemOpen
  \bibfield  {author} {\bibinfo {author} {\bibfnamefont {U.}~\bibnamefont
  {Bhattacharya}}\ and\ \bibinfo {author} {\bibfnamefont {A.}~\bibnamefont
  {Dutta}},\ }\href {\doibase 10.1103/PhysRevB.95.184307} {\bibfield  {journal}
  {\bibinfo  {journal} {Phys. Rev. B}\ }\textbf {\bibinfo {volume} {95}},\
  \bibinfo {pages} {184307} (\bibinfo {year} {2017}{\natexlab{b}})}\BibitemShut
  {NoStop}%
\bibitem [{\citenamefont {Mera}\ \emph {et~al.}(2018)\citenamefont {Mera},
  \citenamefont {Vlachou}, \citenamefont {Paunkovi\ifmmode~\acute{c}\else
  \'{c}\fi{}}, \citenamefont {Vieira},\ and\ \citenamefont
  {Viyuela}}]{mera2017dynamical}%
  \BibitemOpen
  \bibfield  {author} {\bibinfo {author} {\bibfnamefont {B.}~\bibnamefont
  {Mera}}, \bibinfo {author} {\bibfnamefont {C.}~\bibnamefont {Vlachou}},
  \bibinfo {author} {\bibfnamefont {N.}~\bibnamefont
  {Paunkovi\ifmmode~\acute{c}\else \'{c}\fi{}}}, \bibinfo {author}
  {\bibfnamefont {V.~R.}\ \bibnamefont {Vieira}}, \ and\ \bibinfo {author}
  {\bibfnamefont {O.}~\bibnamefont {Viyuela}},\ }\href {\doibase
  10.1103/PhysRevB.97.094110} {\bibfield  {journal} {\bibinfo  {journal} {Phys.
  Rev. B}\ }\textbf {\bibinfo {volume} {97}},\ \bibinfo {pages} {094110}
  (\bibinfo {year} {2018})}\BibitemShut {NoStop}%
\bibitem [{\citenamefont {Landau}\ and\ \citenamefont
  {Lifshitz}(2013)}]{landau2013course}%
  \BibitemOpen
  \bibfield  {author} {\bibinfo {author} {\bibfnamefont {L.~D.}\ \bibnamefont
  {Landau}}\ and\ \bibinfo {author} {\bibfnamefont {E.~M.}\ \bibnamefont
  {Lifshitz}},\ }\href@noop {} {\emph {\bibinfo {title} {{Course of theoretical
  physics}}}}\ (\bibinfo  {publisher} {Elsevier},\ \bibinfo {year}
  {2013})\BibitemShut {NoStop}%
\bibitem [{\citenamefont {Pfeuty}(1970)}]{pfeuty1970one}%
  \BibitemOpen
  \bibfield  {author} {\bibinfo {author} {\bibfnamefont {P.}~\bibnamefont
  {Pfeuty}},\ }\href {\doibase https://doi.org/10.1016/0003-4916(70)90270-8}
  {\bibfield  {journal} {\bibinfo  {journal} {Annals of Physics}\ }\textbf
  {\bibinfo {volume} {57}},\ \bibinfo {pages} {79 } (\bibinfo {year}
  {1970})}\BibitemShut {NoStop}%
\bibitem [{\citenamefont {Pollmann}\ \emph {et~al.}(2010)\citenamefont
  {Pollmann}, \citenamefont {Mukerjee}, \citenamefont {Green},\ and\
  \citenamefont {Moore}}]{pollmann2010}%
  \BibitemOpen
  \bibfield  {author} {\bibinfo {author} {\bibfnamefont {F.}~\bibnamefont
  {Pollmann}}, \bibinfo {author} {\bibfnamefont {S.}~\bibnamefont {Mukerjee}},
  \bibinfo {author} {\bibfnamefont {A.~G.}\ \bibnamefont {Green}}, \ and\
  \bibinfo {author} {\bibfnamefont {J.~E.}\ \bibnamefont {Moore}},\ }\href
  {\doibase 10.1103/PhysRevE.81.020101} {\bibfield  {journal} {\bibinfo
  {journal} {Phys. Rev. E}\ }\textbf {\bibinfo {volume} {81}},\ \bibinfo
  {pages} {020101} (\bibinfo {year} {2010})}\BibitemShut {NoStop}%
\bibitem [{\citenamefont {Vajna}\ and\ \citenamefont
  {D{\'o}ra}(2014)}]{vajna2014}%
  \BibitemOpen
  \bibfield  {author} {\bibinfo {author} {\bibfnamefont {S.}~\bibnamefont
  {Vajna}}\ and\ \bibinfo {author} {\bibfnamefont {B.}~\bibnamefont
  {D{\'o}ra}},\ }\href {\doibase 10.1103/PhysRevB.89.161105} {\bibfield
  {journal} {\bibinfo  {journal} {Phys. Rev. B}\ }\textbf {\bibinfo {volume}
  {89}},\ \bibinfo {pages} {161105} (\bibinfo {year} {2014})}\BibitemShut
  {NoStop}%
\bibitem [{\citenamefont {Abeling}\ and\ \citenamefont
  {Kehrein}(2016)}]{Abeling2016}%
  \BibitemOpen
  \bibfield  {author} {\bibinfo {author} {\bibfnamefont {N.~O.}\ \bibnamefont
  {Abeling}}\ and\ \bibinfo {author} {\bibfnamefont {S.}~\bibnamefont
  {Kehrein}},\ }\href {\doibase 10.1103/PhysRevB.93.104302} {\bibfield
  {journal} {\bibinfo  {journal} {Phys. Rev. B}\ }\textbf {\bibinfo {volume}
  {93}},\ \bibinfo {pages} {104302} (\bibinfo {year} {2016})}\BibitemShut
  {NoStop}%
\bibitem [{\citenamefont {Puskarov}\ and\ \citenamefont
  {Schuricht}(2016)}]{Puskarov2016}%
  \BibitemOpen
  \bibfield  {author} {\bibinfo {author} {\bibfnamefont {T.}~\bibnamefont
  {Puskarov}}\ and\ \bibinfo {author} {\bibfnamefont {D.}~\bibnamefont
  {Schuricht}},\ }\href {https://scipost.org/10.21468/SciPostPhys.1.1.003}
  {\bibfield  {journal} {\bibinfo  {journal} {SciPost Phys.}\ }\textbf
  {\bibinfo {volume} {1}},\ \bibinfo {pages} {003} (\bibinfo {year}
  {2016})}\BibitemShut {NoStop}%
\bibitem [{\citenamefont {Heyl}(2017)}]{Heyl2017critical}%
  \BibitemOpen
  \bibfield  {author} {\bibinfo {author} {\bibfnamefont {M.}~\bibnamefont
  {Heyl}},\ }\href {\doibase 10.1103/PhysRevB.95.060504} {\bibfield  {journal}
  {\bibinfo  {journal} {Phys. Rev. B}\ }\textbf {\bibinfo {volume} {95}},\
  \bibinfo {pages} {060504} (\bibinfo {year} {2017})}\BibitemShut {NoStop}%
\bibitem [{\citenamefont {Bhattacharjee}\ and\ \citenamefont
  {Dutta}(2018)}]{dutta2018}%
  \BibitemOpen
  \bibfield  {author} {\bibinfo {author} {\bibfnamefont {S.}~\bibnamefont
  {Bhattacharjee}}\ and\ \bibinfo {author} {\bibfnamefont {A.}~\bibnamefont
  {Dutta}},\ }\href {\doibase 10.1103/PhysRevB.97.134306} {\bibfield  {journal}
  {\bibinfo  {journal} {Phys. Rev. B}\ }\textbf {\bibinfo {volume} {97}},\
  \bibinfo {pages} {134306} (\bibinfo {year} {2018})}\BibitemShut {NoStop}%
\bibitem [{\citenamefont {\ifmmode \check{Z}\else
  \v{Z}\fi{}unkovi\ifmmode~\check{c}\else \v{c}\fi{}}\ \emph
  {et~al.}(2018)\citenamefont {\ifmmode \check{Z}\else
  \v{Z}\fi{}unkovi\ifmmode~\check{c}\else \v{c}\fi{}}, \citenamefont {Heyl},
  \citenamefont {Knap},\ and\ \citenamefont {Silva}}]{zunkovic2016}%
  \BibitemOpen
  \bibfield  {author} {\bibinfo {author} {\bibfnamefont {B.}~\bibnamefont
  {\ifmmode \check{Z}\else \v{Z}\fi{}unkovi\ifmmode~\check{c}\else
  \v{c}\fi{}}}, \bibinfo {author} {\bibfnamefont {M.}~\bibnamefont {Heyl}},
  \bibinfo {author} {\bibfnamefont {M.}~\bibnamefont {Knap}}, \ and\ \bibinfo
  {author} {\bibfnamefont {A.}~\bibnamefont {Silva}},\ }\href {\doibase
  10.1103/PhysRevLett.120.130601} {\bibfield  {journal} {\bibinfo  {journal}
  {Phys. Rev. Lett.}\ }\textbf {\bibinfo {volume} {120}},\ \bibinfo {pages}
  {130601} (\bibinfo {year} {2018})}\BibitemShut {NoStop}%
\bibitem [{\citenamefont {Zauner-Stauber}\ and\ \citenamefont
  {Halimeh}(2017)}]{zauner2017probing}%
  \BibitemOpen
  \bibfield  {author} {\bibinfo {author} {\bibfnamefont {V.}~\bibnamefont
  {Zauner-Stauber}}\ and\ \bibinfo {author} {\bibfnamefont {J.~C.}\
  \bibnamefont {Halimeh}},\ }\href {\doibase 10.1103/PhysRevE.96.062118}
  {\bibfield  {journal} {\bibinfo  {journal} {Phys. Rev. E}\ }\textbf {\bibinfo
  {volume} {96}},\ \bibinfo {pages} {062118} (\bibinfo {year}
  {2017})}\BibitemShut {NoStop}%
\bibitem [{\citenamefont {Halimeh}\ and\ \citenamefont
  {Zauner-Stauber}(2017)}]{halimeh2017dynamical}%
  \BibitemOpen
  \bibfield  {author} {\bibinfo {author} {\bibfnamefont {J.~C.}\ \bibnamefont
  {Halimeh}}\ and\ \bibinfo {author} {\bibfnamefont {V.}~\bibnamefont
  {Zauner-Stauber}},\ }\href {\doibase 10.1103/PhysRevB.96.134427} {\bibfield
  {journal} {\bibinfo  {journal} {Phys. Rev. B}\ }\textbf {\bibinfo {volume}
  {96}},\ \bibinfo {pages} {134427} (\bibinfo {year} {2017})}\BibitemShut
  {NoStop}%
\bibitem [{\citenamefont {Homrighausen}\ \emph {et~al.}(2017)\citenamefont
  {Homrighausen}, \citenamefont {Abeling}, \citenamefont {Zauner-Stauber},\
  and\ \citenamefont {Halimeh}}]{homrighausen2017}%
  \BibitemOpen
  \bibfield  {author} {\bibinfo {author} {\bibfnamefont {I.}~\bibnamefont
  {Homrighausen}}, \bibinfo {author} {\bibfnamefont {N.~O.}\ \bibnamefont
  {Abeling}}, \bibinfo {author} {\bibfnamefont {V.}~\bibnamefont
  {Zauner-Stauber}}, \ and\ \bibinfo {author} {\bibfnamefont {J.~C.}\
  \bibnamefont {Halimeh}},\ }\href {\doibase 10.1103/PhysRevB.96.104436}
  {\bibfield  {journal} {\bibinfo  {journal} {Phys. Rev. B}\ }\textbf {\bibinfo
  {volume} {96}},\ \bibinfo {pages} {104436} (\bibinfo {year}
  {2017})}\BibitemShut {NoStop}%
\bibitem [{\citenamefont {Lang}\ \emph {et~al.}(2017)\citenamefont {Lang},
  \citenamefont {Frank},\ and\ \citenamefont {Halimeh}}]{lang2018}%
  \BibitemOpen
  \bibfield  {author} {\bibinfo {author} {\bibfnamefont {J.}~\bibnamefont
  {Lang}}, \bibinfo {author} {\bibfnamefont {B.}~\bibnamefont {Frank}}, \ and\
  \bibinfo {author} {\bibfnamefont {J.~C.}\ \bibnamefont {Halimeh}},\ }\href
  {http://arxiv.org/abs/1712.02175v2; http://arxiv.org/pdf/1712.02175v2}
  {\bibfield  {journal} {\bibinfo  {journal} {arXiv: 1712.02175}\ } (\bibinfo
  {year} {2017})}\BibitemShut {NoStop}%
\bibitem [{\citenamefont {Nelson}\ and\ \citenamefont
  {Fisher}(1975)}]{nelson1975}%
  \BibitemOpen
  \bibfield  {author} {\bibinfo {author} {\bibfnamefont {D.~R.}\ \bibnamefont
  {Nelson}}\ and\ \bibinfo {author} {\bibfnamefont {M.~E.}\ \bibnamefont
  {Fisher}},\ }\href {\doibase https://doi.org/10.1016/0003-4916(75)90284-5}
  {\bibfield  {journal} {\bibinfo  {journal} {Annals of Physics}\ }\textbf
  {\bibinfo {volume} {91}},\ \bibinfo {pages} {226 } (\bibinfo {year}
  {1975})}\BibitemShut {NoStop}%
\bibitem [{\citenamefont {Suzuki}(1985)}]{suzuki1985transfer}%
  \BibitemOpen
  \bibfield  {author} {\bibinfo {author} {\bibfnamefont {M.}~\bibnamefont
  {Suzuki}},\ }\href {\doibase 10.1103/PhysRevB.31.2957} {\bibfield  {journal}
  {\bibinfo  {journal} {Phys. Rev. B}\ }\textbf {\bibinfo {volume} {31}},\
  \bibinfo {pages} {2957} (\bibinfo {year} {1985})}\BibitemShut {NoStop}%
\bibitem [{\citenamefont {Andraschko}\ and\ \citenamefont
  {Sirker}(2014)}]{andraschko2014}%
  \BibitemOpen
  \bibfield  {author} {\bibinfo {author} {\bibfnamefont {F.}~\bibnamefont
  {Andraschko}}\ and\ \bibinfo {author} {\bibfnamefont {J.}~\bibnamefont
  {Sirker}},\ }\href {\doibase 10.1103/PhysRevB.89.125120} {\bibfield
  {journal} {\bibinfo  {journal} {Phys. Rev. B}\ }\textbf {\bibinfo {volume}
  {89}},\ \bibinfo {pages} {125120} (\bibinfo {year} {2014})}\BibitemShut
  {NoStop}%
\bibitem [{\citenamefont {Touchette}(2009)}]{touchette2009large}%
  \BibitemOpen
  \bibfield  {author} {\bibinfo {author} {\bibfnamefont {H.}~\bibnamefont
  {Touchette}},\ }\href {\doibase
  https://doi.org/10.1016/j.physrep.2009.05.002} {\bibfield  {journal}
  {\bibinfo  {journal} {Physics Reports}\ }\textbf {\bibinfo {volume} {478}},\
  \bibinfo {pages} {1 } (\bibinfo {year} {2009})}\BibitemShut {NoStop}%
\bibitem [{\citenamefont {Stanley}(1999)}]{stanley1999scaling}%
  \BibitemOpen
  \bibfield  {author} {\bibinfo {author} {\bibfnamefont {H.~E.}\ \bibnamefont
  {Stanley}},\ }\href {\doibase 10.1103/RevModPhys.71.S358} {\bibfield
  {journal} {\bibinfo  {journal} {Rev. Mod. Phys.}\ }\textbf {\bibinfo {volume}
  {71}},\ \bibinfo {pages} {S358} (\bibinfo {year} {1999})}\BibitemShut
  {NoStop}%
\bibitem [{\citenamefont {Bravyi}\ \emph {et~al.}(2011)\citenamefont {Bravyi},
  \citenamefont {DiVincenzo},\ and\ \citenamefont
  {Loss}}]{bravyi2011schrieffer}%
  \BibitemOpen
  \bibfield  {author} {\bibinfo {author} {\bibfnamefont {S.}~\bibnamefont
  {Bravyi}}, \bibinfo {author} {\bibfnamefont {D.~P.}\ \bibnamefont
  {DiVincenzo}}, \ and\ \bibinfo {author} {\bibfnamefont {D.}~\bibnamefont
  {Loss}},\ }\href {\doibase https://doi.org/10.1016/j.aop.2011.06.004}
  {\bibfield  {journal} {\bibinfo  {journal} {Annals of Physics}\ }\textbf
  {\bibinfo {volume} {326}},\ \bibinfo {pages} {2793 } (\bibinfo {year}
  {2011})}\BibitemShut {NoStop}%
\bibitem [{\citenamefont {Heyl}(2014)}]{heyl2014dynamical}%
  \BibitemOpen
  \bibfield  {author} {\bibinfo {author} {\bibfnamefont {M.}~\bibnamefont
  {Heyl}},\ }\href {\doibase 10.1103/PhysRevLett.113.205701} {\bibfield
  {journal} {\bibinfo  {journal} {Phys. Rev. Lett.}\ }\textbf {\bibinfo
  {volume} {113}},\ \bibinfo {pages} {205701} (\bibinfo {year}
  {2014})}\BibitemShut {NoStop}%
\end{thebibliography}%

\end{document}